%% file: main.tex
\documentclass[referee,pdflatex,sn-basic]{sn-jnl}

\hypersetup{hypertexnames=false}

\usepackage{graphicx}
\usepackage{amsmath,amssymb,amsfonts}
\usepackage{amsthm}
\usepackage[title]{appendix}
\usepackage{xcolor}
\usepackage{booktabs}
\usepackage{algorithm}
\usepackage{algorithmicx}
\usepackage{algpseudocode}
\usepackage{tabularx}
\usepackage{array}
\usepackage{enumitem}
\usepackage{url}
\usepackage{acro}

\DeclareAcronym{es}{short=ES, long=engine spread}
\DeclareAcronym{iui}{short=IUI, long=implementation uncertainty interval}
\DeclareAcronym{daf}{short=DAF, long=divergence amplification factor}
\DeclareAcronym{csi}{short=CSI, long=conclusion sensitivity indicator}
\DeclareAcronym{ml}{short=ML, long=machine learning}
\DeclareAcronym{sma}{short=SMA, long=simple moving average}
\DeclareAcronym{tost}{short=TOST, long=two one-sided tests}
\DeclareAcronym{ccc}{short=CCC, long=concordance correlation coefficient}
\DeclareAcronym{aum}{short=AUM, long=assets under management}
\DeclareAcronym{gbr}{short=GBR, long=gradient-boosted regression}
\DeclareAcronym{rf}{short=RF, long=random forest}
\DeclareAcronym{mlp}{short=MLP, long=multi-layer perceptron}
\DeclareAcronym{api}{short=API, long=application programming interface}
\DeclareAcronym{gum}{short=GUM, long=Guide to the Expression of Uncertainty in Measurement}

\graphicspath{{images/}}

\theoremstyle{thmstyleone}
\newtheorem{theorem}{Theorem}

\newtheorem{conjecture}[theorem]{Conjecture}

\theoremstyle{thmstyletwo}

\theoremstyle{thmstylethree}

\newcommand{\ES}{\text{ES}}

\newcommand{\DAF}{\text{DAF}}
\newcommand{\CSI}{\text{CSI}}

\renewcommand{\orcid}[1]{}

\begin{document}

\title[Implementation Risk in Portfolio Backtesting]{Implementation Risk in Portfolio Backtesting: A Previously Unquantified Source of Error}

\author*[1]{\fnm{Don} \sur{Yin}}\email{dy323@cam.ac.uk}\orcid{0000-0002-8971-1057}
\author[2]{\fnm{Takeshi} \sur{Miki}}\email{tm858@cam.ac.uk}\orcid{0009-0000-2063-6274}
\author[3]{\fnm{Vladislav} \sur{Lesnichenko}}\email{lesnichenko.study@gmail.com}\orcid{0009-0004-9023-3613}
\author[4]{\fnm{Vasyl} \sur{Gural}}\email{vasylguraljr@gmail.com}

\affil*[1]{\orgdiv{Department of Clinical Neurosciences}, \orgname{University of Cambridge}, \orgaddress{\city{Cambridge}, \country{United Kingdom}}}
\affil[2]{\orgname{Saltkeep}}
\affil[3,4]{\orgname{Independent Researcher}}

\abstract{Every portfolio backtest depends on a simulation engine, yet no study has asked whether different engines produce the same result for the same strategy. We isolate this question by running 15 benchmark strategies through five independent engines (a purpose-built reference implementation and four open-source libraries) on 30 stratified asset buckets drawn from 180 S\&P~500 constituents over five years under four transaction-cost regimes (0, 18, 36, and 60 basis points). We propose four metrics grounded in metrology and sensitivity analysis to quantify what we term \emph{implementation risk}: engine spread (ES), implementation uncertainty interval (IUI), divergence amplification factor (DAF), and conclusion sensitivity indicator (CSI). At zero transaction cost all engines agree exactly, which establishes that divergence under nonzero costs is attributable to cost-model differences alone. Cost-driven divergence is structured, not random: it scales monotonically with cost intensity (Spearman $\rho = 0.93$, $p < 0.001$), remains below 0.75 percentage points for 12 of 15 benchmarks, but reaches 3.71\% in total return for high-turnover rotation strategies under the heaviest cost regime, an ambiguity equivalent to roughly \$37\,M per year for a \$1\,B portfolio. Engine choice does not reverse the Sharpe-ratio sign in our sample (CSI~$= 0$), but forensic analysis of three engines examined during quality control uncovers seven previously undocumented defects, including a library default in Backtrader that silently divides the user-specified commission rate by 100 before applying it. These findings establish that implementation risk is a material and previously invisible source of error for cost-intensive strategies, and that multi-engine comparison is the most effective diagnostic for detecting it.}



\maketitle
\acresetall

\input{chapters/01-introduction.tex}
\input{chapters/02-related-work.tex}
\input{chapters/03-framework.tex}
\input{chapters/04-benchmark-design.tex}
\input{chapters/05-results.tex}
\input{chapters/06-forensic-analysis.tex}
\input{chapters/07-practical-implications.tex}
\input{chapters/08-conclusion.tex}

\backmatter

\bmhead{Acknowledgements}

The authors thank the developers of bt, vectorbt, backtrader, and cvxportfolio for maintaining open-source backtesting engines that made this study possible.

\section*{Declarations}

\begin{itemize}[nosep]
\item \textbf{Funding:} Not applicable.
\item \textbf{Competing interests:} The authors declare no competing interests.
\item \textbf{Ethics approval:} Not applicable. This study uses publicly available financial market data and involves no human or animal subjects.
\item \textbf{Data availability:} Derived data (metrics, equity curves, divergence tables) and bucket definitions will be deposited at Zenodo upon acceptance. Raw price data can be reproduced via the included download script using Yahoo Finance.
\item \textbf{Code availability:} The benchmark suite, engine wrappers, and our backtesting engine will be released at \url{https://github.com/don-yin/backtest-engine} under the MIT licence upon acceptance.
\item \textbf{Author contribution:} D.Y.\ designed the study, developed the backtesting engine and benchmark suite, conducted the experiments, performed the statistical analysis, and wrote the manuscript. T.M.\ contributed to the experimental design and code review. V.L.\ contributed to the forensic analysis. V.G.\ contributed to the engine integration and testing.
\end{itemize}

\begin{appendices}
\makeatletter
\@addtoreset{figure}{section}
\@addtoreset{table}{section}
\makeatother
\renewcommand{\thefigure}{\Alph{section}\arabic{figure}}
\renewcommand{\thetable}{\Alph{section}\arabic{table}}
\renewcommand{\theHfigure}{app.\Alph{section}.\arabic{figure}}
\renewcommand{\theHtable}{app.\Alph{section}.\arabic{table}}

\input{chapters/A1-algorithm.tex}
\input{chapters/A2-engine-details.tex}
\input{chapters/A3-statistical-diagnostics.tex}
\input{chapters/A4-data-quality.tex}

\end{appendices}

\bibliography{main}

\end{document}

%% file: chapters/01-introduction.tex
\section{Introduction}\label{sec:intro}

Nearly every quantitative finance study relies on a backtest.
Whether the study proposes a new factor, evaluates an allocation rule, or stress-tests a hedging strategy, the backtest is the de facto standard of evidence.
Yet no paper has tested the backtesting engine itself.
The code that simulates order execution, applies transaction costs, marks portfolios to market, and compounds returns is implicitly trusted, treated as though the choice of implementation were inconsequential.
This paper asks a simple question: does it matter which backtesting engine you use?

We introduce the concept of \textit{implementation risk}, defined as the variability in backtest outcomes attributable solely to the choice of simulation engine, with strategy logic, input data, and cost specification held fixed.
In other words, if two research teams implement the identical strategy on the identical data with the identical fee schedule but press ``run'' in different open-source backtesting libraries, how different can their results be?
The answer, we show, depends almost entirely on how the engine handles trading costs, and the magnitude is large enough to matter in practice.

The question is timely for two reasons.
First, the Federal Reserve's SR~11-7 guidance \citep{sr117_2011} establishes a comprehensive framework for model risk management and requires institutions to validate every model that informs a material decision.
Portfolio backtests clearly meet that threshold, yet SR~11-7 says nothing about simulation-engine risk; its examples focus on pricing models, loss-given-default estimators, and stress-test projections rather than on the software that replays historical trades.
Second, a growing literature on backtest overfitting \citep{bailey_deflated_2014} has sharpened awareness that reported Sharpe ratios can be inflated by data mining, multiple testing, and in-sample selection bias.
That literature addresses \textit{statistical} error (the risk of picking a strategy that looked good by chance) but does not address \textit{engine-level} error (the risk that two pieces of code produce different numbers for the same strategy).
Implementation risk is therefore a second, independent source of backtest unreliability that existing safeguards leave unexamined.
To our knowledge, this channel has received no systematic scrutiny despite decades of backtest-driven allocation decisions.

Our approach is designed for causal isolation.
We construct a reusable benchmark suite comprising 15~strategies in five categories: simple allocation rules (equal weight, buy-and-hold, inverse volatility), signal-driven strategies (\ac{sma} momentum, cross-sectional momentum), \ac{ml} signals (four model families), high-turnover rotation strategies that stress-test cost sensitivity, and a zero-cost ablation control.
Each strategy is executed through a reference implementation and four independent open-source backtesting engines on 30~non-overlapping, stratified asset buckets drawn from 180~S\&P~500 constituents over the period 2020--2024.
The reference implementation's cost model is a direct, auditable translation of the proportional-cost specification (Algorithm~\ref{alg:backtest}); divergence under nonzero costs therefore reflects each engine's cost-model implementation choices relative to this specification.
The stratification ensures that results are not artefacts of a single asset universe; the five-year window provides a mix of bull, bear, and sideways regimes including the COVID-19 drawdown and the 2022 rate-hiking cycle.

To quantify divergence we develop a formal measurement framework with four metrics.
\Ac{es} captures the cross-engine coefficient of variation for each performance statistic.
The \ac{iui} provides an uncertainty band around each reported statistic, motivated by measurement uncertainty in metrology.
The \ac{daf} measures how much a cost-bearing benchmark amplifies divergence relative to its zero-cost baseline.
The \ac{csi} flags the most dangerous case: whether the sign of the Sharpe ratio flips across engines, which means one engine says the strategy is profitable while another says it is not.

The results are striking: at zero cost all five engines agree exactly, which indicates that divergence is attributable to the cost model alone.
Under nonzero costs, divergence rises monotonically with cost intensity (Spearman $\rho = 0.93$, $p < 0.001$) yet remains below 0.75\% for 12 of the 15 benchmarks.
Engine choice never reverses the sign of the Sharpe ratio.
Forensic analysis of engines examined during quality control reveals seven defects spanning five failure-mode categories: cost-model bugs, infrastructure bugs, architectural bugs, specification divergence, and complexity sensitivity.

The practical implications are concrete and quantifiable: even modest divergences translate to material dollar-value ambiguity for institutional portfolios (Section~\ref{sec:implications}).
Fund managers, risk officers, and regulators can use our metrics as acceptance criteria: if \ac{es} on a candidate strategy exceeds a tolerable threshold, the backtest should be re-run on a second engine before capital is allocated.

This paper makes three contributions.
First, it provides the first large-scale comparison of backtesting engines, filling a gap in the model-risk literature.
Second, it proposes a formal measurement framework, grounded in metrology and sensitivity-analysis principles, that practitioners can apply to quantify implementation risk in comparable settings.
Third, it releases a fully reusable open benchmark suite, including a purpose-built reference implementation, strategy definitions, data-sampling protocol, and evaluation code, so that future engine developers can certify correctness against a common reference.

The remainder of the paper is organised as follows.
Section~\ref{sec:related} reviews related work on model risk, backtest overfitting, and engine comparisons.
Section~\ref{sec:framework} formalises the implementation-risk framework and defines the four metrics.
Section~\ref{sec:design} describes the experimental design, including the benchmark strategies, engine selection, data pipeline, and cost specifications.
Section~\ref{sec:results} presents the main results.
Section~\ref{sec:forensic} reports the forensic failure-mode analysis.
Section~\ref{sec:implications} discusses practical implications, and Section~\ref{sec:conclusion} concludes.

%% file: chapters/02-related-work.tex
\section{Related Work}\label{sec:related}

Our study sits at the intersection of three literatures: model risk regulation, backtest overfitting, and backtesting-engine design.
We review each in turn and highlight where existing work stops and our contribution begins.

\subsection{Model risk and SR 11-7}\label{sec:related:modelrisk}

The Federal Reserve's SR~11-7 guidance \citep{sr117_2011} defines model risk as ``the potential for adverse consequences from decisions based on incorrect or misused model outputs and reports.''
The guidance requires that every model informing a material business decision undergo independent validation, including assessment of conceptual soundness, outcome analysis, and ongoing monitoring.
Although SR~11-7 is broadly scoped, its illustrative examples (pricing models, allowance-for-loan-loss estimators, stress-test projections) do not mention backtesting engines explicitly.
In practice this means that institutions validate the \textit{strategy} (the model) but not the \textit{simulator} (the engine that produces the backtest numbers the strategy is judged on).

\citet{alonso_model_2022} provide a useful parallel.
They quantify model risk for credit-scoring applications by comparing multiple machine-learning classifiers on the same loan portfolio and measuring dispersion in discriminatory power (the area under the receiver operating characteristic curve).
Their ``models'' correspond to our ``engines''; their dependent variable is classification performance, ours is portfolio return; their regulatory anchor is the internal ratings-based validation framework under the Basel Accords, ours is the Sharpe ratio that triggers an allocation decision.
The analogy is direct: if replacing one credit-scoring model with another changes the capital requirement, that is model risk; if replacing one backtesting engine with another changes the reported Sharpe ratio, that is implementation risk.
\citet{giudici_safe_2023} propose a broader trustworthiness framework for AI in finance and argue that models must be assessed across multiple dimensions (sustainability, accuracy, fairness, explainability) to meet regulatory standards such as the European AI Act.
Our work adds a further dimension, implementation consistency, to the simulation layer of quantitative finance.

\subsection{Backtest overfitting and multiple testing}\label{sec:related:overfitting}

A large and influential literature addresses the statistical reliability of backtests.
\citet{white_reality_2000} introduces the Reality Check, a bootstrap procedure that accounts for the full universe of strategies examined when testing whether the best performer is genuinely superior.
\citet{hansen_test_2005} refines this with the Superior Predictive Ability test, which improves power by re-centring the null distribution.
\citet{romano_stepwise_2005} propose a stepwise extension that controls the family-wise error rate while identifying all strategies that significantly outperform the benchmark, not just the single best.
These methods protect against data-snooping bias: the risk that a strategy appears profitable only because it was selected from a large search space.

\citet{bailey_deflated_2014} sharpen the critique with the deflated Sharpe ratio, which adjusts the significance threshold for a reported Sharpe ratio by the number of trials, skewness, and kurtosis of the return series.
In a companion paper, \citet{bailey_probability_2014} analyse the effects of backtest overfitting on out-of-sample performance and show that the probability of selecting an overfit strategy grows rapidly with the number of trials.
\citet{harvey_and_2016} survey the broader ``factor zoo'' problem and recommend that new factors clear a $t$-statistic hurdle of 3.0 rather than the conventional 2.0, given the hundreds of factors already tested.

Our work is complementary to this literature, not a substitute.
The overfitting literature addresses \textit{which strategy to trust} given that many were tried; we address \textit{which number to trust} given that the same strategy can produce different outputs in different engines.
Statistical overfitting and implementation risk are two independent sources of backtest unreliability: a strategy can pass every multiple-testing correction yet still yield materially different Sharpe ratios depending on the engine that runs it.
Conversely, a strategy can show perfect cross-engine agreement yet still be overfit to the in-sample period.
A complete validation framework must address both.

\subsection{Engine comparisons}\label{sec:related:engines}

Despite the proliferation of open-source backtesting libraries, no academic paper runs the same strategy on the same data through multiple engines and measures the divergence in reported performance.
\citet{wiecki_all_2016} present the closest related work: they compare in-sample and out-of-sample performance of strategies submitted to the Quantopian platform and demonstrate that in-sample Sharpe ratios are poor predictors of live returns.
Their analysis, however, uses a single engine (Zipline); the performance gap they document reflects overfitting and regime change, not engine-level divergence.

\citet{perignon_computational_2024} provide the most direct evidence to date on computational reproducibility in finance.
They ask 168 teams to reproduce the findings of six published papers using the original code and data, and find that only 52\% of 1,000 tests yield the same conclusion.
Their study isolates researcher-level variation (differences in software environments, package versions, and execution paths) rather than engine-level variation (the same algorithm run through different simulation engines on identical inputs).
Our work complements theirs: we hold the researcher constant and vary the engine, which isolates a narrower but previously unmeasured channel of implementation risk.

\citet{boyd_multiperiod_2017} introduce \texttt{cvxportfolio}, a convex-optimisation framework for multi-period trading with realistic transaction costs.
While the library is carefully engineered and includes a sophisticated cost model, the authors benchmark only against themselves (a single-period mean-variance baseline), never against an independent engine.
The same pattern holds across the broader ecosystem: Backtrader, VectorBT, bt, and other popular libraries publish speed benchmarks, feature matrices, and \ac{api} comparisons, but none publish cross-engine correctness comparisons.

The gap is understandable; constructing a fair comparison requires aligning data feeds, rebalancing calendars, cost specifications, and return-computation conventions across libraries that were never designed to interoperate, but the gap is consequential.
Without cross-engine validation, every result produced by a single engine carries an unquantified margin of implementation uncertainty.
This absence is notable given the maturity of the backtesting ecosystem; comparable engineering disciplines (from compiler testing to numerical weather prediction) adopted cross-tool verification decades ago.

%% file: chapters/03-framework.tex
\section{Implementation Risk Framework}\label{sec:framework}

This section formalises the concept of implementation risk and introduces four quantitative metrics for measuring it. it then proposes a taxonomy of failure modes observed in backtesting engines and states a scaling conjecture that links divergence to cost intensity.

\subsection{Formal definition}\label{sec:framework:definition}

Let $s$ denote a fully specified investment strategy (including signal construction, rebalancing rule, and position-sizing logic), $d$ a fixed dataset of asset prices, and $c$ a cost specification (commission rate, slippage model, borrowing cost).
Let $\mathcal{E} = \{e_1, e_2, \dots, e_K\}$ be a set of $K$ backtesting engines, each of which implements a function $R_{e_k}(s, d, c)$ that maps the triple $(s, d, c)$ to a return series.
We define \textit{implementation risk} as the variability of outcomes across engines for fixed inputs:
\begin{equation}\label{eq:impl_risk}
  \text{IR}(s, d, c) \;=\; \operatorname{Var}_{e \in \mathcal{E}}\!\bigl[\,f\!\bigl(R_e(s, d, c)\bigr)\bigr],
\end{equation}
where $f(\cdot)$ is a summary statistic of interest (e.g.\ annualised Sharpe ratio, cumulative return, maximum drawdown).
When $\text{IR} = 0$ for all summary statistics (Equation~\ref{eq:impl_risk}), the engines are \textit{implementation-equivalent} for that $(s, d, c)$ triple.

Conceptually, we treat the $K$ engines as exchangeable draws from a population of plausible implementations, analogous to the treatment of measurement instruments in ISO~5725 \citep{iso5725_1994}, where multiple laboratories measure the same quantity and the inter-laboratory variance quantifies reproducibility.
A design choice is that $s$, $d$, and $c$ must be held strictly constant across engines; any residual variation is then attributable to implementation differences alone, which enables causal isolation.

\subsection{Metrics}\label{sec:framework:metrics}

We propose four complementary metrics, each of which captures a different facet of implementation risk.
Together they move from magnitude (how large is the spread?) to sensitivity (does the spread change conclusions?).

The concept of software-induced variability has precedents in other domains: climate science calls it structural uncertainty or inter-model spread, metrology calls it inter-laboratory variability (ISO~5725), and software engineering calls it numerical reproducibility.
Each of our four metrics adapts a tool from one of these traditions to the backtesting context; the methodological novelty lies in the domain-specific synthesis rather than in the invention of new statistical machinery.

\subsubsection{\texorpdfstring{\Acf{es}}{Engine spread (ES)}}
Engine spread quantifies the relative dispersion of a summary statistic across engines.
For a given strategy-bucket-cost combination, let $\{f_1, \dots, f_K\}$ be the values of summary statistic $f$ produced by $K$ engines.
The coefficient-of-variation form is
\begin{equation}\label{eq:es_cv}
  \text{ES}_f^{\text{cv}} \;=\; \frac{\sigma_f}{\bar{f}} \times 100,
\end{equation}
where $\bar{f}$ and $\sigma_f$ are the cross-engine mean and standard deviation.
The range form, useful when the distribution is heavy-tailed or $K$ is small, is
\begin{equation}\label{eq:es_range}
  \text{ES}_f^{\text{range}} \;=\; \max_k f_k \;-\; \min_k f_k.
\end{equation}
Both variants are reported in our results.
In practice we report \acs{es}$^{\text{range}}$ (Equation~\ref{eq:es_range}) as the primary metric throughout the results section because the coefficient-of-variation form (Equation~\ref{eq:es_cv}) becomes unstable when the cross-engine mean $\bar{f}$ is near zero.
The concept mirrors the ``multi-model spread'' used in climate-projection ensembles by \citet{tebaldi_use_2007}, where the same forcing scenario is run through multiple general circulation models and the inter-model spread characterises structural uncertainty.

\subsubsection{\texorpdfstring{\Acf{iui}}{Implementation uncertainty interval (IUI)}}
While engine spread summarises dispersion in a single number, practitioners often need an interval estimate: a band around the reported Sharpe ratio within which the ``true'' engine-agnostic value plausibly lies.
We define:
\begin{equation}\label{eq:iui}
  \text{IUI}_f \;=\; \bar{f} \;\pm\; t_{0.025,\,K-1}\;\cdot\;s_f,
\end{equation}
where $s_f$ is the sample standard deviation across $K$ engines (with $K - 1$ degrees of freedom) and $t_{0.025,\,K-1}$ is the Student-$t$ critical value.
The $t$-multiplier accounts for the uncertainty in $s_f$ when $K$ is small (here $K = 5$, so $t_{0.025,4} \approx 2.78$).
The construction is motivated by the expanded-uncertainty framework of the \acf{gum} \citep{jcgm_gum_2008}, but differs from \acs{gum} in one respect: \acs{gum} targets the uncertainty of a mean estimate (using $s_f / \sqrt{K}$), whereas \acs{iui} uses $s_f$ directly to characterise the full spread of engine outputs.
The resulting band is wider than a confidence interval for the population mean and narrower than a formal $(p, \gamma)$ tolerance interval; it serves as a practical uncertainty band for inter-engine variability rather than a formally calibrated statistical interval.
A narrow \acs{iui} means the reported statistic is robust to engine choice; a wide \acs{iui} signals that the backtest result is engine-dependent and warrants scrutiny.

The exchangeability assumption underlying \acs{iui} is a modelling convenience, not a claim about random sampling.
The five engines were selected because they are the most widely adopted open-source Python backtesting libraries, not drawn from a well-defined population.
The \acs{iui} therefore captures a variability band conditional on this specific engine sample; adding engines from outside the current sample could widen or narrow it.

\subsubsection{\texorpdfstring{\Acf{daf}}{Divergence amplification factor (DAF)}}
Different experimental conditions (cost levels, rebalancing frequencies, strategy complexity) may amplify or attenuate implementation risk.
The divergence amplification factor captures this by comparing the engine spread of a given benchmark to that of a minimal-complexity reference benchmark (in our experiments, BM01, equal-weight monthly rebalancing):
\begin{equation}\label{eq:daf}
  \text{DAF}_f \;=\; \frac{\text{ES}_f^{\text{range}}(\text{benchmark})}{\text{ES}_f^{\text{range}}(\text{reference})}.
\end{equation}
\Acs{daf} is well defined whenever the reference benchmark exhibits nonzero engine spread; BM01 satisfies this condition with $\text{ES}^{\text{range}} = 0.379$~percentage points (pp).
Conceptually, \acs{daf} is analogous to a condition number in numerical analysis: it measures how much a perturbation in the ``input'' (engine choice) is amplified in the ``output'' (performance statistic) as the problem difficulty (cost intensity) increases.
The framing draws on the sensitivity-analysis tradition of \citet{saltelli_global_2008}, who advocate decomposing output variance into contributions from individual input factors.

\subsubsection{\texorpdfstring{\Acf{csi}}{Conclusion sensitivity indicator (CSI)}}
The most consequential form of implementation risk is not a large spread per se but a qualitative reversal: one engine reports a positive Sharpe ratio (strategy is profitable) while another reports a negative one (strategy destroys value).
The conclusion sensitivity indicator is a binary flag:
\begin{equation}\label{eq:csi}
  \text{CSI}(s, d, c) \;=\;
  \begin{cases}
    1 & \text{if } \exists\; j, k \text{ such that } \operatorname{sgn}(f_j) \neq \operatorname{sgn}(f_k), \\
    0 & \text{otherwise},
  \end{cases}
\end{equation}
where $f_j$ and $f_k$ are the Sharpe ratios from engines $j$ and $k$.
A $\text{CSI} = 1$ is an immediate red flag: the investment decision (go/no-go) depends on which engine was used, precisely the kind of outcome SR~11-7 \citep{sr117_2011} warns against.
The indicator is inspired by the concept of \textit{predictive multiplicity} introduced by \citet{marx_predictive_2020}, who show that competing models can assign conflicting predictions to the same individual; \acs{csi} transposes this idea from classification to portfolio evaluation.

\subsection{Failure-mode taxonomy}\label{sec:framework:taxonomy}

Forensic analysis of engines excluded during our quality-control phase reveals five recurring categories of implementation failure.
We define each briefly; detailed case studies appear in Section~\ref{sec:forensic}.

\begin{enumerate}
  \item \textbf{Cost-model bugs.} The engine applies transaction costs incorrectly, for example, it may charge commission on notional value rather than on the traded delta, or double-count slippage on round-trip trades. This is the dominant source of divergence in our sample.

  \item \textbf{Infrastructure bugs.} Errors in calendar alignment, price look-up, or dividend handling cause the engine to use stale or misaligned data when it computes portfolio values. These bugs produce small but persistent drift.

  \item \textbf{Architectural bugs.} Limitations inherent in the engine's design, such as inability to process fractional shares, forced end-of-day settlement when the strategy specifies intraday execution, or hard-coded assumptions about trading frequency, constrain the strategy specification and introduce systematic bias.

  \item \textbf{Specification divergence.} The engine interprets an ambiguous specification differently from other engines, for example, it may apply costs at trade time versus at settlement time, or compute returns on a gross versus net basis. Neither interpretation is ``wrong'' in isolation, but the inconsistency across engines creates divergence.

  \item \textbf{Complexity sensitivity.} The engine handles simple strategies correctly but introduces errors as strategy complexity increases (more assets, higher turnover, non-linear cost models), often because internal numerical routines lose precision or buffer sizes overflow.
\end{enumerate}

\subsection{A scaling conjecture}\label{sec:framework:proposition}

The observation that divergence scales with cost intensity admits a simple formalisation.

\begin{conjecture}\label{conj:divergence_scaling}
  Let $N$ denote the number of rebalancing events over the backtest horizon and $\delta_c$ the per-trade cost-model discrepancy between two engines.
  If cost errors accumulate approximately linearly across trades, the cumulative return divergence $D$ satisfies $D = \mathcal{O}(N \cdot \delta_c)$.
\end{conjecture}

The intuition is straightforward: each rebalance introduces a cost-model error of order $\delta_c$, and these errors accumulate over $N$ trades.
Strategies with high turnover (large $N$) or complex cost models (large $\delta_c$) should therefore exhibit greater implementation risk.
We test this prediction in Section~\ref{sec:results}, where the rank correlation between a composite cost-intensity score and engine spread is $\rho = 0.93$.
we regard this as the simplest non-trivial prediction the framework generates; the strength of its observed support suggests that cost-model disagreement is the dominant channel.
The linear scaling is an approximation; in practice, compounding effects, path-dependent cost models, and margin constraints can produce super-linear accumulation, but the first-order term dominates in the moderate-turnover regime covered by our benchmark suite.

%% file: chapters/04-benchmark-design.tex
\section{Benchmark Design}\label{sec:design}

This section describes the experimental apparatus: strategy suite, asset universe, engine selection, cost specification, and statistical analysis plan.
Every component is defined so that an independent team can reproduce the experiment from the released code and data.
The design is intentionally reusable; future studies can plug into the same infrastructure.

\subsection{Strategy suite}

\begin{figure}[htbp]
\centering
\includegraphics[width=\textwidth]{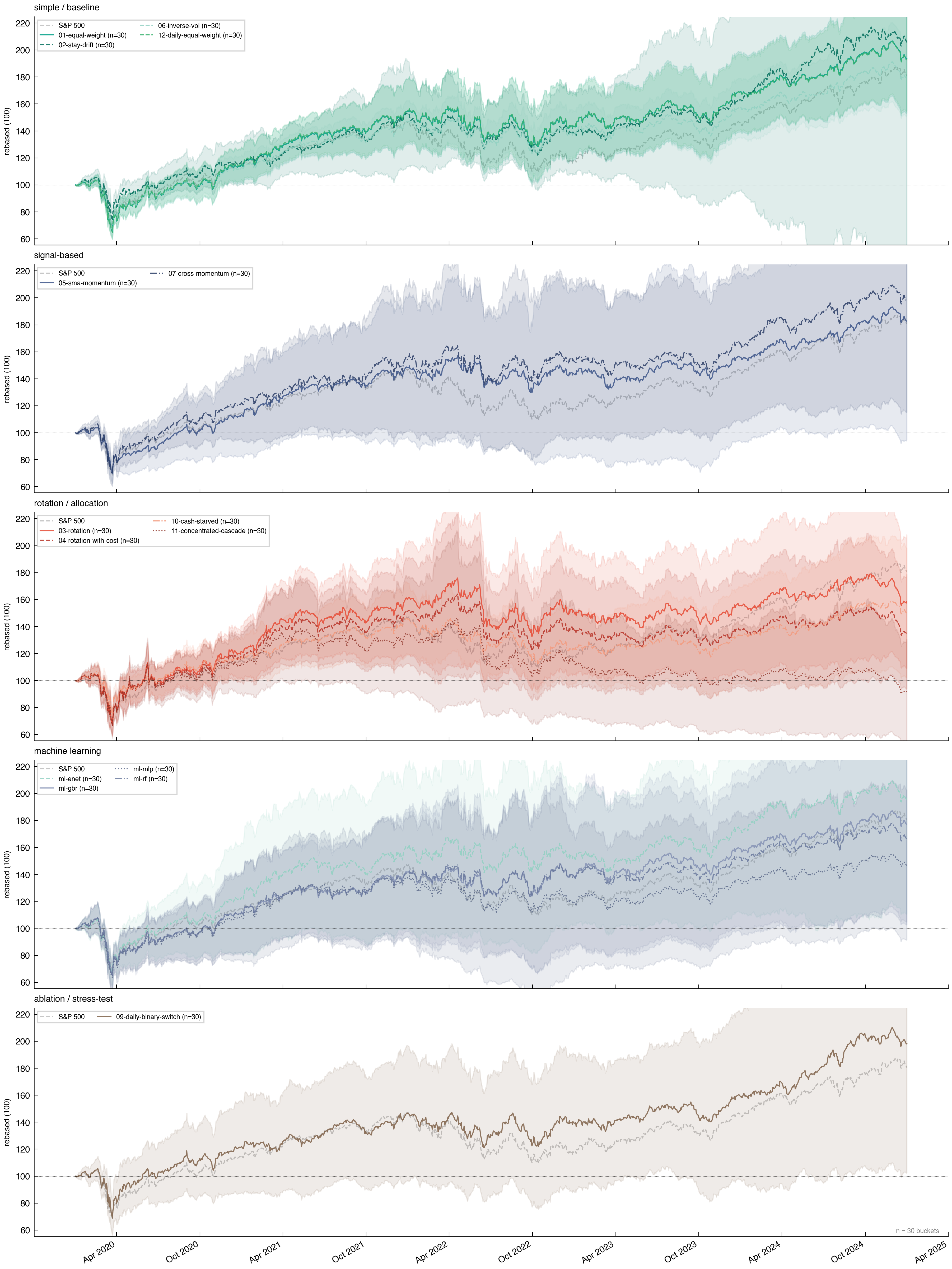}
\caption{Mean equity curves across 30 asset buckets for all 15 benchmark strategies. Shaded bands show $\pm 1$ standard deviation of bucket-to-bucket variation. The five engine traces per benchmark are nearly indistinguishable at this scale, consistent with high agreement for most strategies.}\label{fig:equity}
\end{figure}

We construct 15 benchmark strategies in five categories that span a broad range of portfolio construction methods, turnover profiles, and cost sensitivities (Table~\ref{tab:benchmarks} and Figure~\ref{fig:equity}).

The \textit{simple} category contains four strategies that require no predictive signal.
BM01 applies equal weighting with monthly rebalancing and serves as the canonical baseline.
BM02 is a stay-drift (buy-and-hold) strategy that purchases equal weights on day one and never rebalances, which provides a near-zero-turnover reference.
BM06 weights each asset inversely proportional to its trailing 60-day realised volatility and rebalances monthly.
BM12 replicates BM01 at daily frequency, which increases the number of rebalancing events by a factor of roughly 20 and thereby amplifies any per-trade cost-model discrepancy.

The \textit{signal} category includes two momentum strategies.
BM05 applies a 200-day \acs{sma} filter: the signal is computed daily but trades execute only at monthly rebalance dates; each asset is held when its closing price exceeds the \acs{sma} and sold otherwise.
BM07 implements cross-sectional momentum with a 12-month look-back that excludes the most recent month (12-1), which selects the top two performers from the asset bucket and rebalances monthly.

The \acs{ml} category uses a single template, BM08, instantiated with four learners: \ac{gbr}, \ac{rf}, \ac{mlp}, and elastic net.
All variants share an identical walk-forward protocol with a rolling six-month training window, a 21-day gap between training and prediction, and five features: 21-, 63-, and 126-day returns plus 20- and 60-day realised volatility.
Each month the top two assets by predicted return are selected to test whether engine divergence is amplified when the signal carries estimation noise.

The \textit{rotation/allocation} category stresses turnover and cost sensitivity.
BM03 performs a large rotation that reshuffles the entire portfolio at every monthly rebalance.
BM04 is identical to BM03 but applies a doubled cost multiplier ($2\times$ baseline), which isolates the marginal effect of cost intensity while turnover is held fixed.
BM10 simulates a cash-starved settlement environment in which available capital is allocated in three tiers (60/30/10) to a rotating set of assets, which introduces path-dependent cash constraints.
BM11 is a concentrated cascade that allocates 95\% of the portfolio to a single asset with heavy transaction costs (60~basis points (bps)), which creates the most extreme cost-sensitivity scenario in the suite.

The \textit{ablation} category contains a single strategy, BM09, that serves as the causal isolation control.
BM09 switches daily between two assets at zero transaction cost.
Because no cost model is invoked, any non-zero divergence across engines would indicate a flaw in the experimental harness rather than a cost-model difference.
This strategy is the linchpin of our identification argument: if engines agree exactly at zero cost, all divergence observed under non-zero costs can be attributed to the cost model alone.

\begin{table}[t]
\caption{Benchmark strategy suite. Fifteen strategies in five categories that cover simple allocation rules, signal-driven strategies, machine-learning signals, high-turnover rotation, and a zero-cost ablation control. Cost regimes range from 0 to 60~bps per trade.}\label{tab:benchmarks}
\centering
\begin{tabular}{llllr}
\toprule
ID & Name & Category & Cost regime & Rebal. \\
\midrule
BM01 & equal-weight monthly         & simple    & baseline (18~bps) & monthly \\
BM02 & stay-drift (buy-and-hold)    & simple    & baseline (18~bps) & once    \\
BM03 & large rotation               & rotation  & baseline (18~bps) & monthly \\
BM04 & large rotation + 2$\times$ costs & rotation & 2$\times$ baseline (36~bps) & monthly \\
BM05 & \acs{sma} momentum (200-day)       & signal    & baseline (18~bps) & monthly \\
BM06 & inverse-volatility (60d)     & simple    & baseline (18~bps) & monthly \\
BM07 & cross-momentum (12-1, top~2) & signal    & baseline (18~bps) & monthly \\
BM08$_\text{gbr}$ & \acs{ml} signal, \acs{gbr}          & \acs{ml} & baseline (18~bps) & monthly \\
BM08$_\text{rf}$ & \acs{ml} signal, \acs{rf}           & \acs{ml} & baseline (18~bps) & monthly \\
BM08$_\text{mlp}$ & \acs{ml} signal, \acs{mlp}          & \acs{ml} & baseline (18~bps) & monthly \\
BM08$_\text{enet}$ & \acs{ml} signal, elastic net  & \acs{ml} & baseline (18~bps) & monthly \\
BM09 & daily binary switch           & ablation  & zero (0~bps)      & daily   \\
BM10 & cash-starved settlement       & rotation  & baseline (18~bps) & monthly \\
BM11 & concentrated cascade          & rotation  & heavy (60~bps)    & monthly \\
BM12 & daily equal-weight            & simple    & baseline (18~bps) & daily   \\
\botrule
\end{tabular}
\end{table}

\subsection{Asset universe and stratified buckets}

The asset universe comprises 180 constituents of the S\&P~500, drawn from all 11 Global Industry Classification Standard sectors, with 16--17 stocks per sector.
From these 180 stocks we construct 30 non-overlapping buckets of six stocks each.
Every stock appears in exactly one bucket and every bucket contains six distinct sectors, which ensures that no bucket is dominated by a single industry.

Bucket generation follows the rerandomisation framework of \citet{morgan_rerandomization_2012}, itself rooted in the principles of controlled experimentation introduced by \citet{fisher_design_1935}.
We generate 20{,}000 candidate partitions and retain the one that minimises the Mahalanobis distance across three covariates: annualised volatility, average pairwise correlation, and log total return over the sample period.
The resulting allocation achieves a sector chi-squared statistic of 0.16 ($p = 1.000$) and an entropy ratio of 0.9998, which indicates near-perfect sector balance.

The full downloaded dataset spans January~2018 to December~2024 (1{,}761 trading days).
The first two years (2018--2019) serve exclusively as a look-back buffer for strategies that require historical features: the four \acs{ml} benchmarks (BM08 variants) train on this period, while signal strategies (BM05, BM06, BM07) use it for initial moving-average and momentum warm-up.
All portfolio evaluation takes place over the backtest period of January~2020 to December~2024, which yields 1{,}258 trading days.
All strategies use only daily adjusted-close prices; no open, high, low, or volume data enter any strategy or engine.
Prices were obtained from Yahoo Finance via the \texttt{yfinance} Python library \citep{aroussi_yfinance_2024}.
The adjusted-close series accounts for stock splits and dividend reinvestments, which produces a total-return series suitable for portfolio simulation.
All 180 stocks have complete price histories over this window with zero missing values, which eliminates survivorship-bias and missing-value imputation concerns.
Raw price data are not redistributed; the released code includes a download script that reproduces the exact dataset from Yahoo Finance's public servers.
The backtest period encompasses the COVID-19 drawdown (February--March~2020), the subsequent recovery rally, the 2022 rate-hiking sell-off, and the 2023--2024 rebound, which provides a mix of bull, bear, and range-bound regimes.

\subsection{Engine selection}

Eight backtesting engines were evaluated (Table~\ref{tab:engine-roster}): bt \citep{morissette_bt_2014}, vectorbt \citep{polakow_vectorbt_2020}, Backtrader \citep{rodriguez_backtrader_2015}, cvxportfolio \citep{boyd_multiperiod_2017}, Zipline-Reloaded \citep{jansen_zipline_2020}, NautilusTrader \citep{nautechsystems_nautilus_2021}, QuantConnect Lean \citep{quantconnect_lean_2013}, and our reference implementation.
Five form the active comparison set and three were excluded during quality control.
All five retained engines are pip-installable Python packages, which ensures replicability in any standard Python environment.

\begin{table}[t]
\caption{Backtesting engines evaluated in this study. Five engines form the active comparison set. Zipline-Reloaded was excluded due to an unfixable trading-calendar alignment bug; NautilusTrader was excluded due to four compounding errors (silent truncation, double commissions, position-mode mismatch, and arithmetic overflow); QuantConnect Lean was excluded because its C\# core lacks a native Python \acs{api}. Version numbers refer to the releases used throughout the experiment.}\label{tab:engine-roster}
\centering\footnotesize
\begin{tabular}{@{}llllll@{}}
\toprule
Engine & Package & Version & Language & Architecture & Status \\
\midrule
Ours                & (internal)                & --     & Python      & vectorised      & retained \\
bt                  & \texttt{bt}               & 1.1.0  & Python      & tree-structured & retained \\
vectorbt            & \texttt{vectorbt}         & 0.26.2 & Python      & vectorised      & retained \\
Backtrader          & \texttt{backtrader}       & 1.9.78 & Python      & event-driven    & retained \\
cvxportfolio        & \texttt{cvxportfolio}     & 1.3.0  & Python      & optimisation    & retained \\
\addlinespace
Zipline-Reloaded    & \texttt{zipline-reloaded} & 3.0    & Python      & event-driven    & excluded \\
NautilusTrader      & \texttt{nautilus\_trader} & 1.178  & Rust/Python & event-driven    & excluded \\
QuantConnect Lean   & \texttt{Lean}             & --     & C\#         & event-driven    & excluded \\
\botrule
\end{tabular}
\end{table}

Our reference implementation was developed independently for this study and serves as the ground-truth calibrator.
We acknowledge that this makes the reference correct by construction relative to its own specification; divergence measured against it reflects disagreement with this specification, not absolute error in any external sense.
The zero-cost ablation (BM09) provides an engine-independent validation: all five engines agree exactly when no cost model is invoked, which indicates that the non-cost logic path is equivalent across all implementations.
Its cost computation is an arithmetic translation of Algorithm~\ref{alg:backtest} with no hidden state, no implicit rounding, and no framework abstractions between the specification and the code.

Three additional engines were evaluated but excluded after quality control.
Zipline-Reloaded was excluded due to an unfixable trading-calendar bug that caused date misalignment on rebalancing days.
NautilusTrader exhibited four compounding errors: a position-mode mismatch (HEDGING versus NETTING semantics), silent truncation of the backtest after 62 of 1{,}258 trading days, double-charged commissions, and a fixed-point arithmetic overflow at scale; these could not be resolved within the study period.
QuantConnect Lean was excluded because its C\# core lacks a native Python \acs{api}, which makes identical strategy logic impossible without a confounding translation layer.

\subsection{Transaction cost specification}

All engines apply a proportional cost model in which every trade incurs a fixed fraction of the trade's notional value.
Four cost regimes are used, summarised below:

\begin{center}
\begin{tabular}{lrl}
\toprule
Regime & Cost & Benchmarks \\
\midrule
zero-cost      & 0~bps  & BM09 \\
baseline       & 18~bps & BM01--03, 05--08, 10, 12 \\
2$\times$ baseline & 36~bps & BM04 \\
heavy          & 60~bps & BM11 \\
\botrule
\end{tabular}
\end{center}

The baseline regime of 18~bps decomposes into 15~bps for commission and 3~bps for price impact (slippage), calibrated to the all-in cost of a retail brokerage that executes US large-cap equities in 2024.
The heavy regime of 60~bps (50~bps commission plus 10~bps slippage) is deliberately conservative; for reference, \citet{demiguel_optimal_2009} use 50~bps one-way in their comparison of portfolio rules.
The zero-cost regime applied to BM09 is not a modelling convenience but a methodological necessity: it provides the causal counterfactual against which all cost-driven divergence is measured.

\subsection{Statistical analysis plan}

The analysis plan was specified in full before any engine outputs were compared.\footnote{The analysis plan was formalised in internal project documentation prior to running comparisons but was not deposited with a public pre-registration registry.}
Because every benchmark is run on each of the $\binom{5}{2} = 10$ pairwise engine combinations across 30 independent buckets, the primary unit of observation is the per-bucket pairwise divergence in total return (and, separately, in Sharpe ratio, compound annual growth rate, annualised volatility, and maximum drawdown).

The \textit{primary test} is a one-sample $t$-test of whether the mean pairwise divergence differs from zero, with $p$-values corrected for multiplicity using the Benjamini--Hochberg false-discovery-rate procedure.
This procedure controls the expected proportion of false discoveries at rate $q$ under the positive regression dependence on a subset condition, which \citet{benjamini_control_2001} show is satisfied when test statistics are positively correlated, a property that holds naturally here because the same engine pair contributes to multiple benchmark comparisons.
In total the correction covers 150 primary tests (15 benchmarks $\times$ 10 engine pairs) at $q = 0.05$.

Three \textit{robustness checks} supplement the primary test.
The Wilcoxon signed-rank test provides a non-parametric alternative that does not assume normality.
A Monte Carlo permutation test with 10{,}000 random draws supplies an exact $p$-value free of distributional assumptions.
The \ac{tost} procedure of \citet{schuirmann_comparison_1987} tests for practical equivalence at two pre-specified margins, 10~bps and 50~bps, and addresses the dual question: are divergences not merely statistically significant but also economically meaningful?

\textit{Diagnostic tests} include the Shapiro--Wilk test and quantile--quantile (Q--Q) plots for assessing the normality assumption required by the primary $t$-test.
A pseudo-replication check assesses the exchangeability of the 30 buckets via cluster bootstrap (resampling buckets with replacement and recomputing the cost--divergence Spearman correlation), supplemented by lag-1 autocorrelation of the per-bucket divergence series under lexicographic bucket ordering.

\textit{Concordance analysis} uses Lin's \ac{ccc} \citep{lin_concordance_1989}, computed per benchmark across the 30 buckets, to assess whether engines preserve not only the mean level of each metric but also the within-bucket ranking.
A \acs{ccc} of 1.0 indicates perfect agreement in both location and scale; values below 1.0 reveal systematic biases or rank reversals that a simple mean comparison might miss.

%% file: chapters/05-results.tex
\section{Results}\label{sec:results}

This section presents the findings in five parts: the zero-cost agreement baseline, cost-driven divergence across all 15~benchmarks, ordinal versus cardinal agreement, the four implementation-risk metrics, and a battery of statistical robustness checks.

\subsection{Zero-cost perfect agreement}

BM09, the daily binary switch executed at zero transaction cost, produces identical results across all five engines.
Every one of the $\binom{5}{2} = 10$ pairwise comparisons yields a relative difference of exactly 0.0000\% in total return.
This result validates the experimental harness: when the cost model is removed, all five engines are functionally identical in rebalancing calendar, price look-up convention, and return-compounding arithmetic.
The zero-cost baseline serves as a causal isolation device; any divergence observed under non-zero costs can be attributed exclusively to differences in transaction-cost implementation.
this result is central to the study's identification strategy: it transforms every subsequent divergence measurement from a correlation into a causally identified quantity.

\subsection{Cost-driven divergence}

Table~\ref{tab:divergence} reports the mean and maximum pairwise relative difference in total return for each of the 15~benchmarks, computed across all 10 engine pairs and averaged over 30 asset buckets.

\begin{table}[t]
\caption{Pairwise relative difference in total return (\%) across five engines for each benchmark. Mean and max are taken over the 10 engine pairs after averaging across 30 buckets. Three tiers emerge: simple strategies show a maximum pairwise divergence near 0.18\%, signal and \acs{ml} strategies span 0.28--0.49\%, and rotation strategies reach 3.71\%.}\label{tab:divergence}
\centering
\begin{tabular}{llrrr}
\toprule
ID & Category & Cost (bps) & Mean (\%) & Max (\%) \\
\midrule
BM09 & ablation  &  0 & 0.0000 & 0.0000 \\
BM01 & simple    & 18 & 0.1144 & 0.1799 \\
BM02 & simple    & 18 & 0.1187 & 0.1799 \\
BM06 & simple    & 18 & 0.1136 & 0.1799 \\
BM12 & simple    & 18 & 0.1098 & 0.1799 \\
BM10 & rotation  & 18 & 0.1754 & 0.2720 \\
BM05 & signal    & 18 & 0.1859 & 0.2750 \\
BM07 & signal    & 18 & 0.1999 & 0.3012 \\
BM08$_\text{mlp}$  & \acs{ml} & 18 & 0.2769 & 0.4075 \\
BM08$_\text{enet}$ & \acs{ml} & 18 & 0.2524 & 0.4109 \\
BM08$_\text{gbr}$  & \acs{ml} & 18 & 0.3000 & 0.4672 \\
BM08$_\text{rf}$   & \acs{ml} & 18 & 0.3014 & 0.4881 \\
BM11 & rotation  & 60 & 1.2195 & 2.1552 \\
BM03 & rotation  & 18 & 2.3032 & 3.6091 \\
BM04 & rotation  & 36 & 2.4411 & 3.7077 \\
\botrule
\end{tabular}
\end{table}

\begin{figure}[htbp]
\centering
\includegraphics[width=\textwidth]{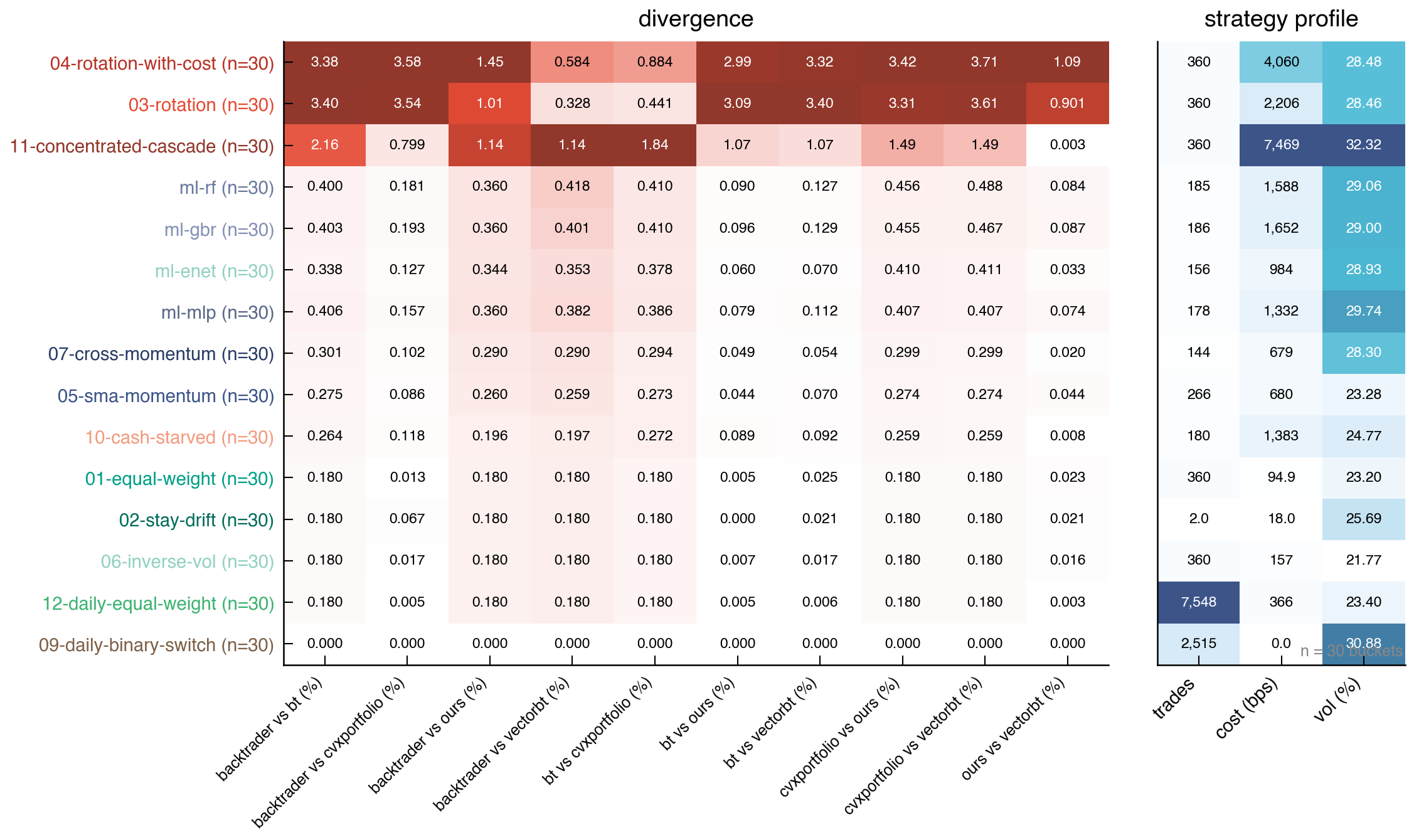}
\caption{Divergence heatmap across all 15 benchmarks and 10 engine pairs. Colour intensity encodes the mean pairwise relative difference in total return; the three-tier structure is clearly visible, with rotation strategies (BM03, BM04, BM11) producing the largest divergences.}\label{fig:divergence-landscape}
\end{figure}

The results fall into three tiers (Figure~\ref{fig:divergence-landscape}).
The first tier comprises the simple and ablation strategies (BM01, BM02, BM06, BM09, BM12), for which the maximum pairwise divergence never exceeds 0.1799\%.
The second tier includes the signal and \acs{ml} strategies (BM05, BM07, BM08 variants), where maximum divergences range from 0.2750\% to 0.4881\%.
The third tier contains the rotation strategies (BM03, BM04, BM10, BM11), which span 0.2720\% (BM10, constrained by tiered cash allocation) to 3.7077\% (BM04, full rotation at doubled cost).

Two patterns stand out.
First, turnover matters more than cost level within the third tier: BM03 and BM04 (full monthly rotation) exhibit maximum divergences of 3.6091\% and 3.7077\%, which exceed BM11 (2.1552\%) even though BM11 carries a heavier per-trade cost of 60~bps.
BM11 concentrates 95\% of capital in a single asset, which limits the total number of cost-bearing trades.
Second, BM10 achieves a low maximum divergence (0.2720\%) despite being a rotation strategy because its tiered cash-allocation rule constrains each trade's notional value.
The near-equality of BM03 (3.6091\%) and BM04 (3.7077\%) despite a doubled cost rate is consistent with the floor decomposition in Section~\ref{sec:results:floor}: the dominant source of divergence is a fixed cost-deduction timing convention (the 0.18\% floor), not the cost rate itself; doubling the rate amplifies only the residual above the floor.

\begin{figure}[htbp]
\centering
\includegraphics[width=\textwidth]{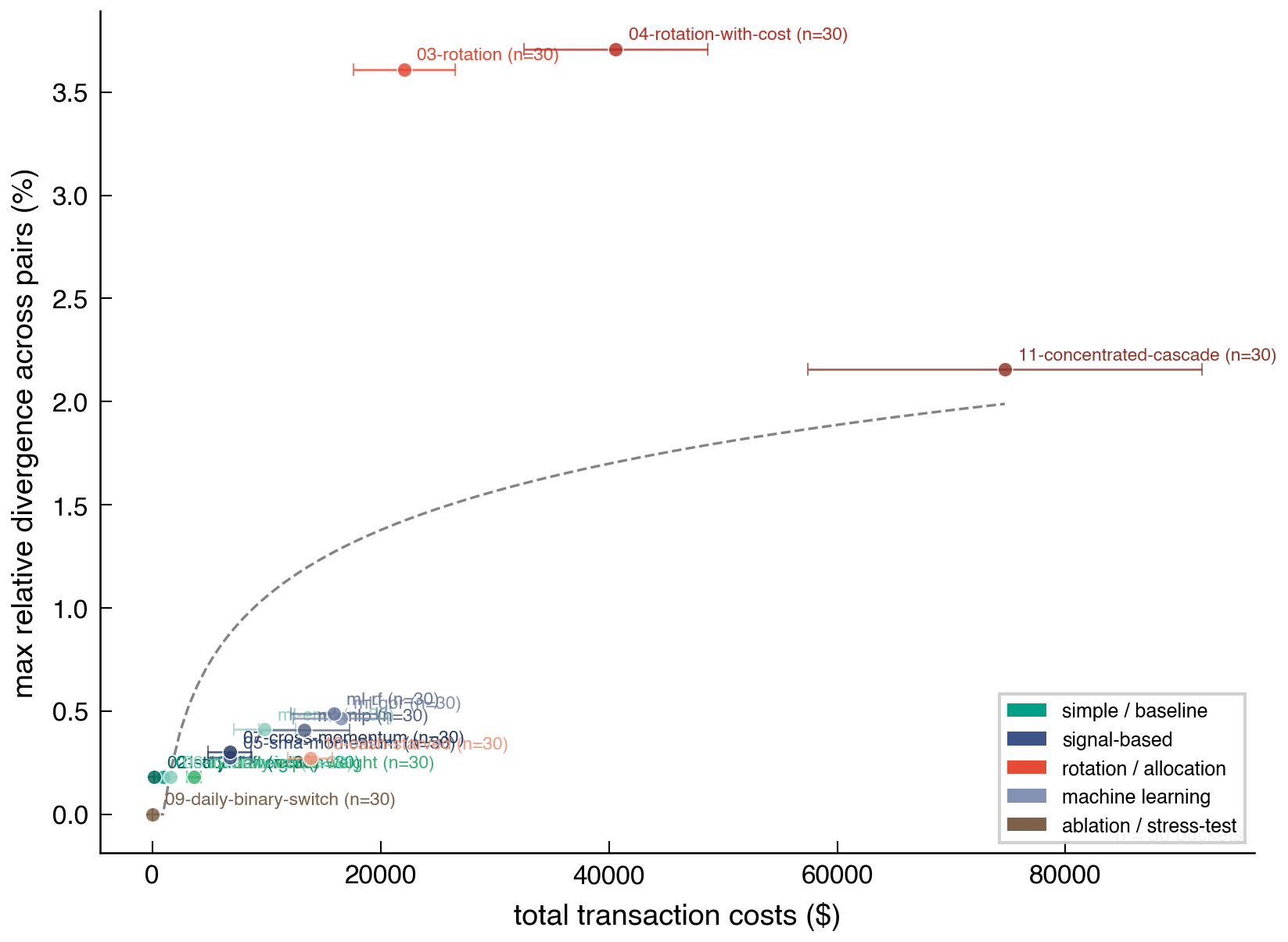}
\caption{Mean pairwise divergence versus composite cost-intensity score. The near-monotonic gradient indicates that cost intensity is the primary predictor of engine disagreement (Spearman $\rho = 0.93$, $p < 0.001$).}\label{fig:divergence-vs-cost}
\end{figure}

The monotonic relationship between cost intensity and divergence is supported quantitatively (Figure~\ref{fig:divergence-vs-cost}).
A composite cost-intensity score, defined as the product of per-trade cost and annual turnover, correlates with the \acs{es} metric: Pearson $r = 0.68$ ($p = 0.006$), Spearman $\rho = 0.93$ ($p < 0.001$), consistent with Conjecture~\ref{conj:divergence_scaling}.

\begin{figure}[htbp]
\centering
\includegraphics[width=\textwidth]{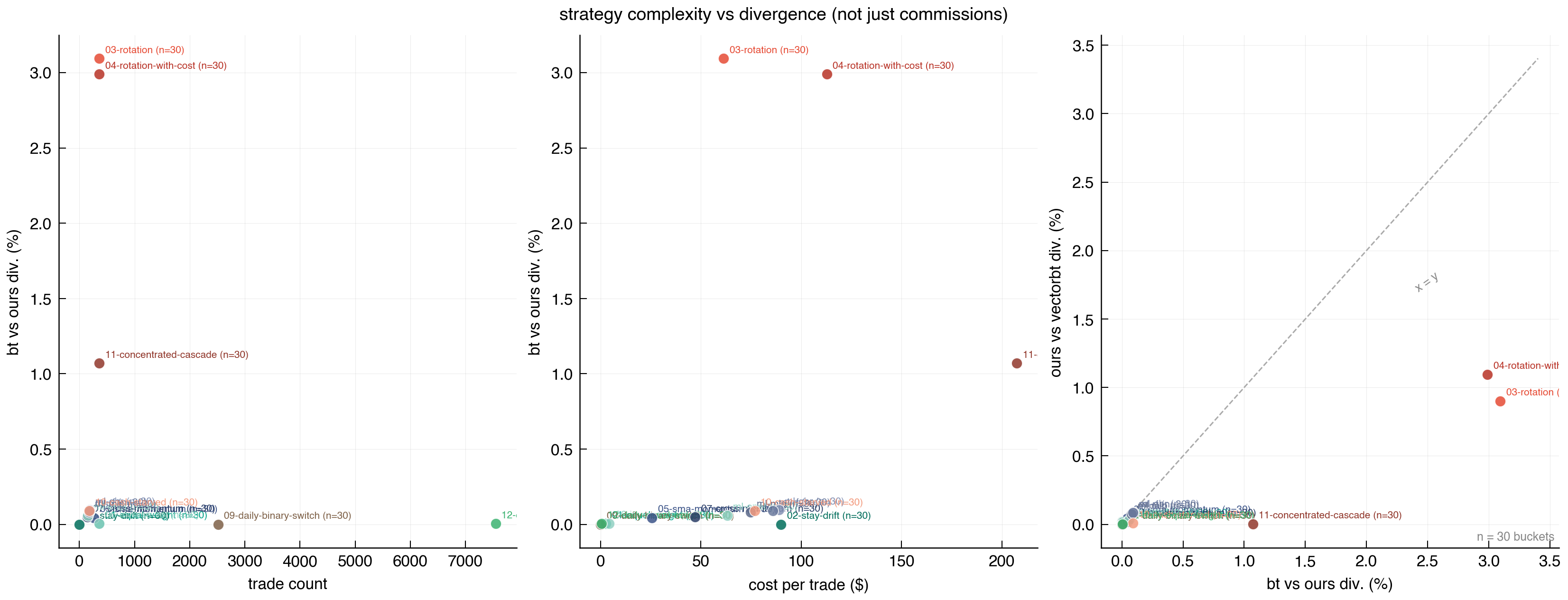}
\caption{Divergence decomposed by strategy complexity. Higher-turnover strategies accumulate more cost-model disagreement, consistent with the linear scaling predicted by Conjecture~\ref{conj:divergence_scaling}.}\label{fig:complexity-analysis}
\end{figure}

Figure~\ref{fig:complexity-analysis} decomposes the cost-intensity relationship by strategy complexity, which provides direct support for Conjecture~\ref{conj:divergence_scaling}.

\begin{figure}[htbp]
\centering
\includegraphics[width=\textwidth]{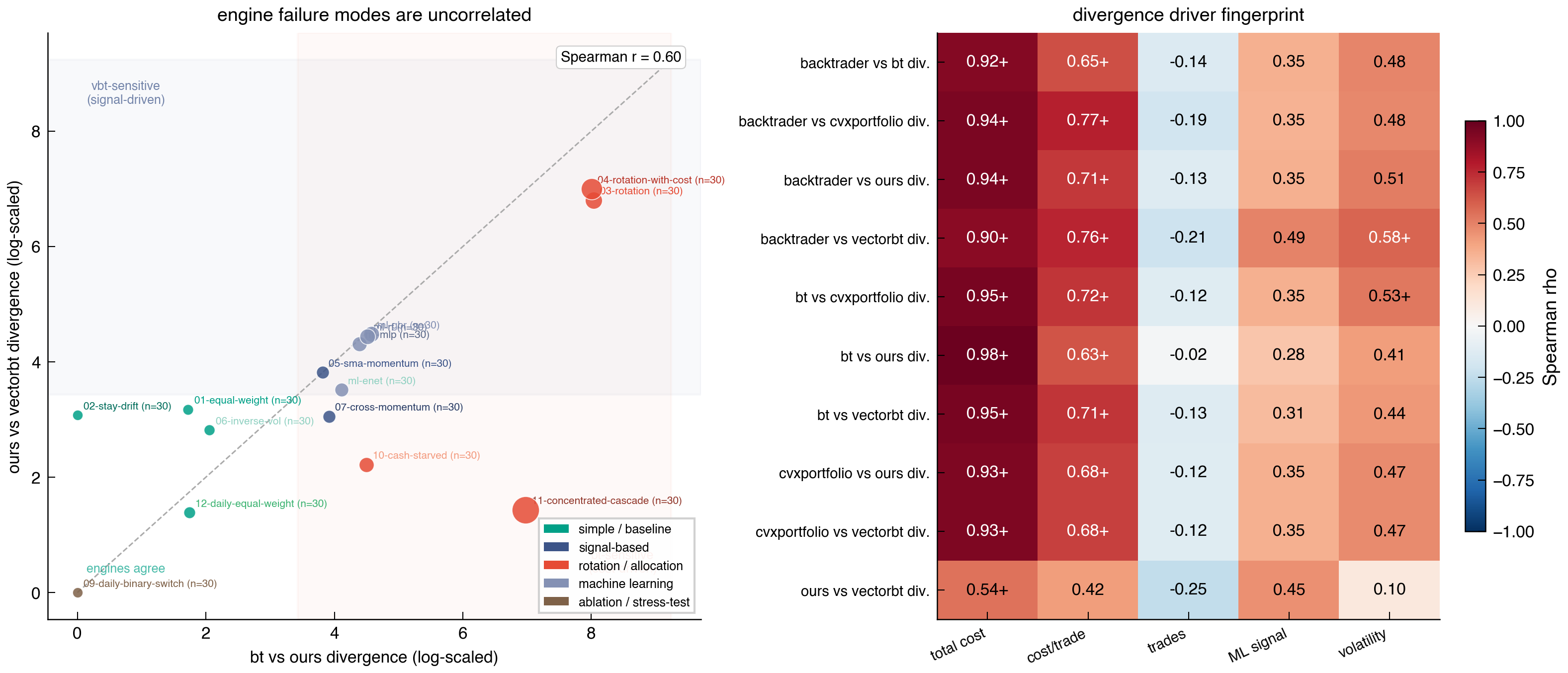}
\caption{Anatomy of inter-engine divergence. Left: per-benchmark divergence for two representative engine pairs on log-scaled axes; the moderate Spearman correlation ($\rho = 0.60$) indicates that pairwise disagreements are only weakly coupled. Right: divergence driver fingerprint showing Spearman correlations between each driver (total cost, cost per trade, trade count, \acs{ml} signal, volatility) and per-benchmark divergence for all ten engine pairs. The heterogeneous fingerprint across pairs indicates that no single engine pair captures the full spectrum of cost-model disagreement.}\label{fig:divergence-anatomy}
\end{figure}

the cost-intensity gradient identifies how much divergence to expect, but does not reveal whether different engine pairs disagree for the same reasons. Figure~\ref{fig:divergence-anatomy} addresses this question. the left panel plots per-benchmark divergence for two representative engine pairs (bt versus ours, and ours versus vectorbt) on log-scaled axes; the Spearman correlation of $\rho = 0.60$ indicates that the magnitude of disagreement for one pair is only weakly predictive of the magnitude for another. the right panel profiles divergence against five driver categories (total cost, cost per trade, trade count, \acs{ml} signal, and volatility) for each of the ten engine pairs via Spearman correlations. the fingerprint varies across pairs: some show stronger sensitivity to cost per trade, while others are more influenced by volatility or \acs{ml} signal composition. this heterogeneity means that validation against a single alternative engine can detect only the subset of disagreements to which that specific pair is sensitive.

\subsection{Ordinal versus cardinal agreement}

Lin's \acs{ccc}, computed per benchmark across the 30 asset buckets, assesses whether engines agree on the within-bucket ranking, not only the mean level.

\begin{figure}[htbp]
\centering
\includegraphics[width=\textwidth]{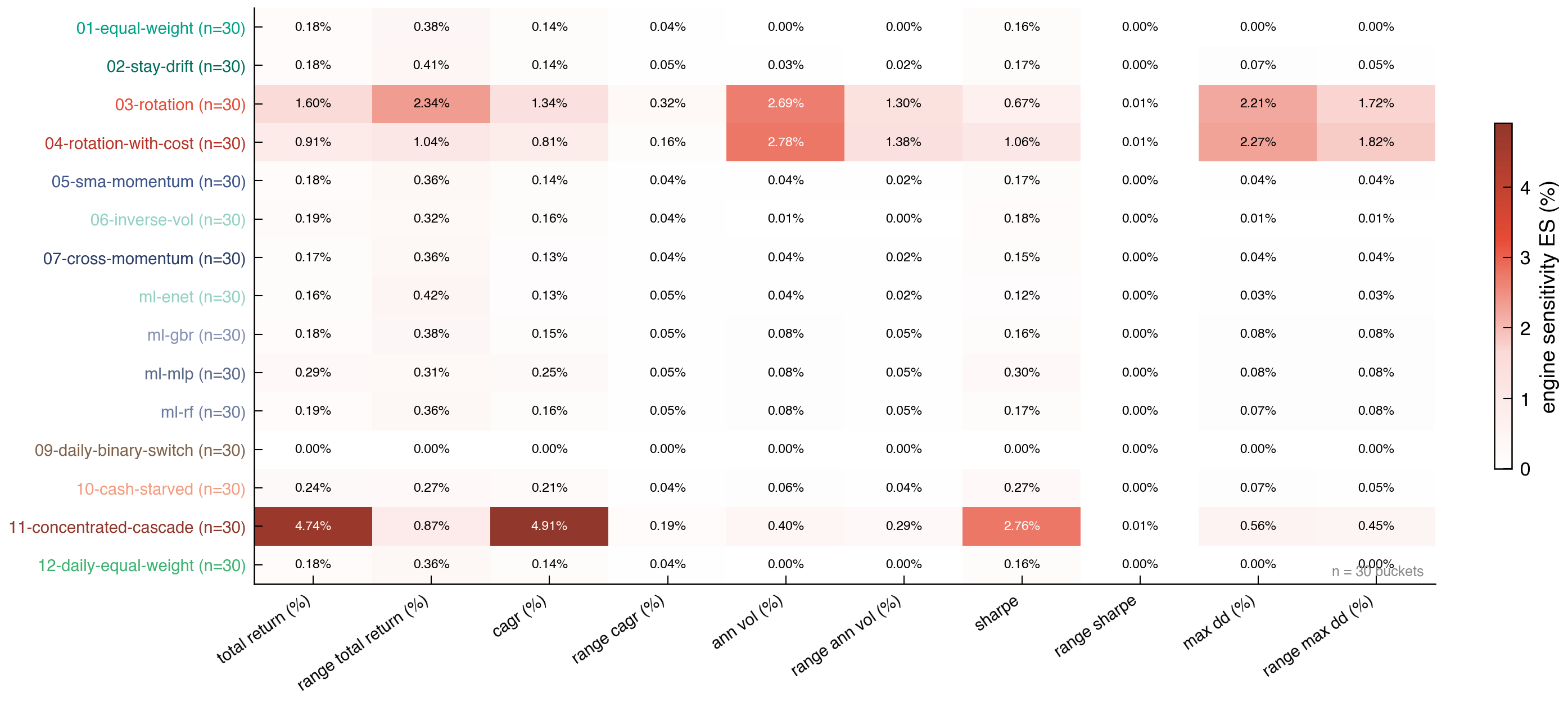}
\caption{Lin's CCC for each benchmark and metric pair. Twelve of fifteen benchmarks show near-perfect concordance ($\geq 1.000$ to three decimal places); the three rotation strategies (BM03, BM04, BM11) show slightly reduced but still strong ordinal agreement.}\label{fig:engine-concordance}
\end{figure}

Twelve of the 15~benchmarks achieve a minimum \acs{ccc} of 1.000 (to three decimal places) on every metric (Figure~\ref{fig:engine-concordance}).
The three exceptions are BM03 ($\text{CCC}_{\text{min,ann\_vol}} = 0.8756$), BM04 ($\text{CCC}_{\text{min,ann\_vol}} = 0.8632$), and BM11 ($\text{CCC}_{\text{min,ann\_vol}} = 0.9958$).
Even in the worst case (BM04), the engines strongly agree on which buckets perform well and which perform poorly.

A researcher using any of the five engines would rank the same asset universes in nearly the same order, even where reported returns differ by several percentage points.

\subsection{Implementation risk metrics}

We report the four metrics from Section~\ref{sec:framework}: \acs{es} (Equation~\ref{eq:es_range}), \acs{iui} (Equation~\ref{eq:iui}), \acs{daf} (Equation~\ref{eq:daf}), and \acs{csi} (Equation~\ref{eq:csi}).

\subsubsection{\texorpdfstring{\acs{es}}{ES}}
\acs{es} (range of total returns across five engines) scales with cost intensity: $\ES = 0.379$~pp for BM01, $\ES = 0.874$~pp for BM11, $\ES = 2.339$~pp for BM03, and $\ES = 1.040$~pp for BM04.

\subsubsection{\texorpdfstring{\acs{iui}}{IUI}}
The \acs{iui} width (uncertainty band) is 1.069~pp for BM01, 2.408~pp for BM11, 5.942~pp for BM03, and 2.079~pp for BM04.
These widths indicate the range of values a practitioner might observe by switching engines; they should accompany any single-engine backtest result as a measure of implementation variability.

\subsubsection{\texorpdfstring{\acs{daf}}{DAF}}
\acs{daf} is computed relative to BM01 (the simplest cost-bearing benchmark, $\text{ES}^{\text{range}} = 0.379$~pp); values greater than one indicate that a benchmark amplifies cost-model disagreement beyond the baseline.
The largest values are BM03 ($\DAF = 6.178$), BM04 ($\DAF = 2.747$), and BM11 ($\DAF = 2.308$); rotation strategies amplify cost-model disagreement by a factor of two to six.

\subsubsection{\texorpdfstring{\acs{csi}}{CSI}}
Across all 15~benchmarks, $\CSI = 0$; engine choice never reverses the sign of the Sharpe ratio.
BM11, the only benchmark with a negative mean Sharpe, illustrates this: all five engines report negative ratios (ours $= -0.1151$, bt $= -0.1219$, vectorbt $= -0.1151$, backtrader $= -0.1225$, cvxportfolio $= -0.1211$), and $\CSI_{\text{sharpe\_frac}} = 1.0$ indicates unanimous agreement on unprofitability.
The closest any benchmark comes to a sign reversal is BM11, where the most negative Sharpe ratio across engines is $-0.1225$ (backtrader) and the least negative is $-0.1151$ (ours and vectorbt).
The gap of 0.0073 in absolute Sharpe ratio means that divergence would need to grow by roughly an order of magnitude beyond the observed range before \acs{csi} could reach one.

\subsection{Statistical robustness}

\subsubsection{\texorpdfstring{\acs{tost}}{TOST} equivalence}
\begin{figure}[htbp]
\centering
\includegraphics[width=\textwidth]{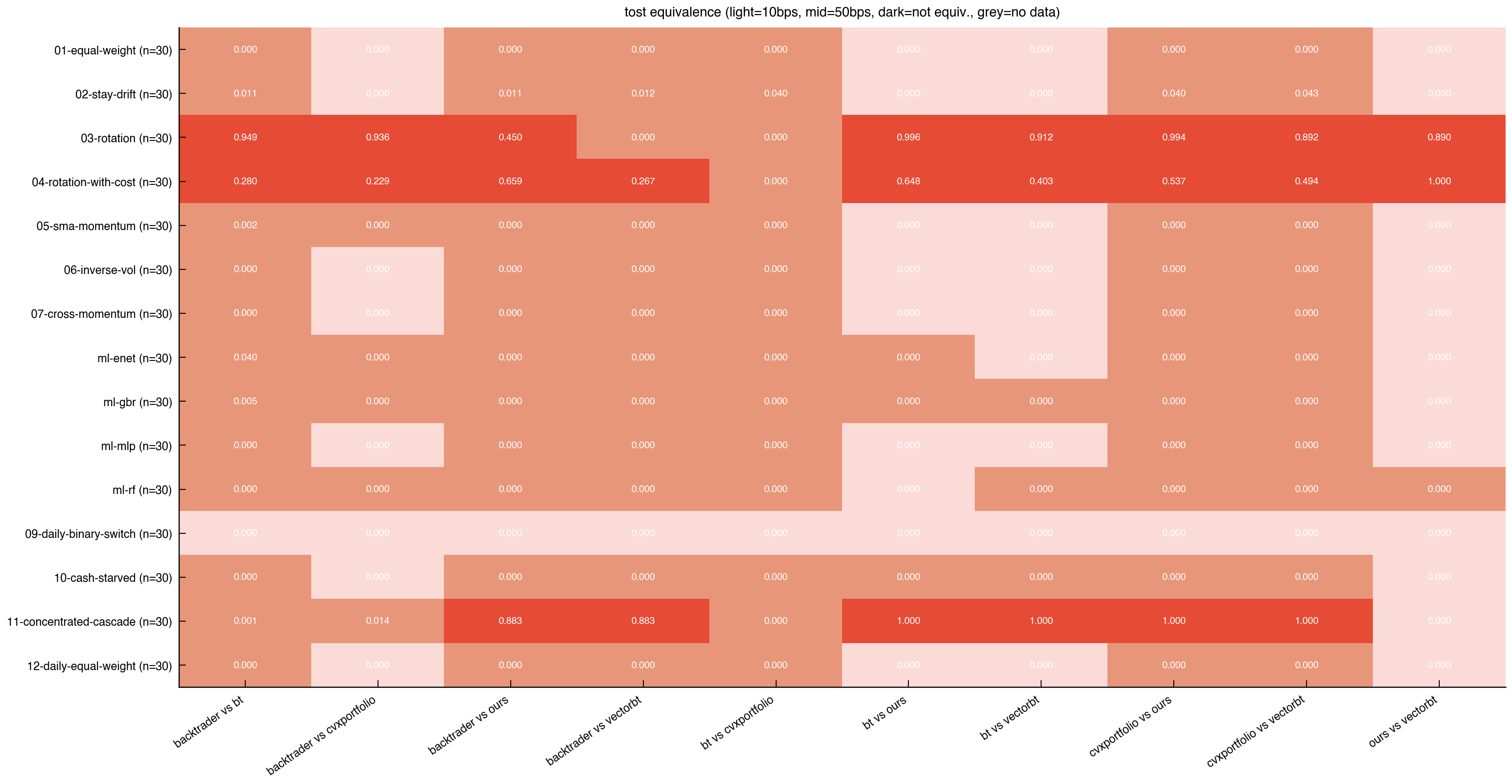}
\caption{\acs{tost} equivalence classification at 10~bps and 50~bps margins across benchmarks and engine pairs. Light cells indicate equivalence at the tighter margin; progressively darker shading indicates equivalence only at 50~bps or non-equivalence. Rotation strategies (BM03, BM04) and BM11 contain the only non-equivalent pairs.}\label{fig:tost-equivalence}
\end{figure}

BM09 achieves equivalence at the 10~bps margin for all 10 engine pairs.
Simple strategies (BM01, BM02, BM06, BM12) achieve equivalence at 50~bps or 10~bps for all pairs.
The only benchmarks with non-equivalent pairs are BM03 and BM04 (many pairs) and BM11 (some pairs), as shown in Figure~\ref{fig:tost-equivalence}.

\subsubsection{Wilcoxon signed-rank test}
The non-parametric Wilcoxon signed-rank test agrees with the parametric $t$-test in every case (Figure~\ref{fig:wilcoxon-robustness} in Appendix~\ref{sec:diagnostics}), which rules out the possibility that conclusions are artefacts of distributional assumptions.

\subsubsection{Permutation test}
A sign-flip permutation test (10{,}000 draws) is applied to each strategy category.
The ablation category (BM09) is excluded because its per-bucket divergences are negligible in magnitude (they round to 0.0000\% at display precision); applying the sign-flip test to floating-point noise would produce a spurious rejection.
For the remaining categories: simple $p = 0.0001$, signal $p = 0.0001$, \acs{ml} $p = 0.0001$, rotation $p = 0.4187$ (not significant).
The non-significance of the rotation category reflects high cross-bucket variance rather than absence of divergence.

\subsubsection{Pseudo-replication diagnostic}
Because all 30 buckets share the same time period (2020--2024) and draw from correlated S\&P~500 stocks, we assess exchangeability via two complementary checks.
First, a cluster bootstrap resamples the 30 buckets with replacement (5{,}000 draws), recomputes the benchmark-level mean divergence, and re-estimates the Spearman $\rho$ between cost intensity and engine spread.
The bootstrap 95\% confidence interval captures sampling variability under potential bucket dependence; values reported in Appendix~\ref{sec:diagnostics}.
Second, lag-1 autocorrelation of the per-bucket divergence series is computed under lexicographic bucket ordering as a supplementary diagnostic.
Per-category means range from $-0.020$ to $0.049$, with 95\% confidence intervals that include or lie close to zero.
We note that lag-1 autocorrelation is sensitive to the choice of ordering; the lexicographic ordering has no structural significance.
The non-overlapping bucket design and rerandomisation balance (Section~\ref{sec:design}) eliminate shared-stock confounds and balance covariates across buckets by construction.
Residual dependence from the shared time period and common market factors may reduce the effective sample size for within-benchmark inference, but the benchmark-level cost--divergence relationship ($\rho = 0.93$) and the three-tier structure are ordinal findings that do not depend on within-bucket independence.

\subsubsection{Floor decomposition}\label{sec:results:floor}
A systematic 0.18\% floor appears across virtually all engine pairs, which traces to an initial-value convention: backtrader and cvxportfolio report pre-trade equity, while our engine, bt, and vectorbt report post-trade equity.
The difference equals the cost of initial portfolio construction at 18~bps on a fully invested portfolio.
For simple strategies the residual above the floor is near zero; for rotation strategies the residual dominates and reflects genuine cost-model disagreement on subsequent trades.

\subsubsection{Power analysis}
a post hoc power analysis (Appendix~\ref{sec:diagnostics}) shows that $n = 30$ buckets provides greater than 80\% power at equivalence margins of 10--25~bps for all non-rotation categories.

%% file: chapters/06-forensic-analysis.tex
\section{Forensic Analysis}\label{sec:forensic}

The preceding section established that divergence among the five retained engines is small, structured, and attributable to cost-model differences.
This section examines implementation defects discovered during engine integration.
Two of the three engines that failed quality control, Zipline-Reloaded and NautilusTrader, are analysed alongside Backtrader, which was retained after the defects documented below were diagnosed and patched.
QuantConnect Lean, the third excluded engine, is omitted from this analysis because its exclusion reflects a language-barrier limitation (C\# core with no native Python \acs{api}) rather than an implementation defect amenable to source-code forensics.
We document these defects in detail for two reasons: they validate the failure-mode taxonomy introduced in Section~\ref{sec:framework}, and they illustrate the kinds of implementation errors that a multi-engine comparison is designed to catch.

\subsection{Backtrader: cost-model misconfiguration}\label{sec:forensic:bt}

Backtrader exhibited two bugs, both definitive upon source-code inspection.

The first is a $100\times$ undercharge of proportional commissions.
In \texttt{CommInfoBase.\_\_init\_\_} (line~157 of \texttt{comminfo.py}), the default parameter \texttt{percabs} is set to \texttt{False}.
When \texttt{percabs=False}, the engine divides the user-supplied commission rate by~100 before applying it.
If the user passes \texttt{commission=0.0018} with the intent of specifying 18~basis points, the engine applies only 0.18~basis points, a factor-of-100 error.
The behaviour is documented in a cryptic docstring but contradicts the convention of every other engine in our sample, all of which interpret the commission rate as an absolute proportion.
The fix is to set \texttt{percabs=True}, which disables the implicit division.
Under our taxonomy this is a \textit{cost-model bug}: the engine's default parameterisation silently misinterprets the user's cost specification.

The second bug concerns fill ordering and margin rejection.
The method \texttt{check\_submitted()} (lines~558--583 of \texttt{bbroker.py}) iterates over pending orders by popping them from a first-in-first-out deque and pseudo-executing each against the running cash balance.
If the running cash balance falls below zero at any point during this sequential pass, the engine triggers an immediate margin rejection.
The consequence is that buy orders whose funding depends on proceeds from sell orders later in the deque are rejected before those sell orders are processed.
This is entirely undocumented.
In a rebalancing context where multiple positions are simultaneously increased and decreased, the order in which the engine processes fills determines whether the portfolio can be constructed at all.
The fix requires atomic delta computation with a sells-first ordering so that proceeds are available before purchases are attempted.
Under our taxonomy this is an \textit{infrastructure bug}: the execution-sequencing logic produces incorrect rejections that are unrelated to the strategy or cost specification.

After both fixes are applied, Backtrader's pairwise divergence against the retained engines falls to 0.00--3.58\%, consistent with the cost-driven divergence structure described in Section~\ref{sec:results}.
The most striking confirmation is benchmark~BM09 (zero cost), which originally showed 97.08\% divergence and drops to exactly 0.000\% after correction, indicating that the anomaly was entirely attributable to the two bugs above.
this collapse from 97\% to exactly zero constitutes a strong form of bug-fix validation and exposes a central danger of single-engine work: before correction, Backtrader produced a plausible equity curve that gave no indication of a hundredfold cost undercharge.

\subsection{Zipline-Reloaded: calendar architecture failure}\label{sec:forensic:zipline}

Zipline-Reloaded failed at the data-ingestion stage with a session-alignment assertion error.
The root cause lies in \texttt{calendar\_utils.get\_calendar}, which delegates to \texttt{exchange\_calendars.get\_calendar} with inconsistent keyword arguments across code paths.
During data ingestion the call passes \texttt{start=1990-01-01}; during runtime the same function is called without the \texttt{start} parameter.
The \texttt{exchange\_calendars} library caches calendar instances by a composite key that includes both the exchange name and the keyword arguments.
Because the two call sites supply different kwargs, the cache produces two distinct calendar instances rather than returning the same one.
Downstream code assumes a single shared calendar and raises an assertion failure when session indices do not align.

This bug cannot be fixed by user-side configuration; it requires a patch to the library's internal calendar-resolution logic to ensure that all call sites pass identical kwargs.
Under our taxonomy this is an \textit{architectural bug}: the failure arises not from a single misparameterised line but from a structural inconsistency in the calendar subsystem's caching contract.

\subsection{NautilusTrader: compounding failures}\label{sec:forensic:nautilus}

NautilusTrader could not be brought into agreement with the other engines despite successive attempts to resolve each issue as it appeared.

The first obstacle was a \textit{specification divergence}.
In its default HEDGING mode the engine generates a new position identifier for every order, so selling shares to reduce a long position creates a separate short position rather than reducing the existing long.
The HEDGING mode is designed for strategies that maintain simultaneous long and short positions in the same instrument, but our benchmarks assume net-position (NETTING) semantics.

Switching to NETTING mode resolved the position-accounting issue but exposed a silent truncation: the engine processed 62 of the 1{,}258 trading days in the sample and then terminated without error or diagnostic output.
Under our taxonomy this is an \textit{infrastructure bug}; the resulting equity curve simply stopped, and the user would have no indication that the backtest was incomplete.

With the truncation worked around, a \textit{cost-model bug} became apparent: commissions were applied twice.
Our engine wrapper reserves cash for expected transaction fees before submitting orders, and the NautilusTrader engine also charges fees internally via its \texttt{MakerTakerFeeModel}.
The combined effect is a double deduction of transaction costs, an integration-layer error in which the wrapper and the engine both assume responsibility for cost accounting.

A further limitation emerged at scale: the engine's fixed-point arithmetic overflows for large notional values, which terminates the process rather than returning an error.
Under our taxonomy this is a \textit{complexity-sensitivity} defect that restricts the engine to modest portfolio sizes.

\subsection{Lessons for the taxonomy}\label{sec:forensic:lessons}

The issues documented above span all five failure-mode categories introduced in Section~\ref{sec:framework}: cost-model bugs (Backtrader's percabs default, NautilusTrader's double commission), infrastructure bugs (Backtrader's fill ordering, NautilusTrader's silent truncation), an architectural bug (Zipline-Reloaded's calendar caching), specification divergence (NautilusTrader's HEDGING semantics), and complexity sensitivity (NautilusTrader's fixed-point overflow).

Cost-model bugs and infrastructure bugs are the most frequent categories; each accounts for two of the seven issues.
This is consistent with the main finding of the paper: divergence among retained engines is driven almost entirely by differences in cost-model implementation.
The forensic cases show that these differences can be far more severe when an engine's default configuration silently misinterprets the user's intent.

More broadly, the fact that all five taxonomy categories appear across just three engines' initial integration failures suggests that the taxonomy is comprehensive, at least for the class of proportional-cost, market-order backtests studied here.
The forensic exercise also shows that multi-engine comparison is not merely a statistical convenience; it is the most efficient diagnostic tool for detecting implementation errors that would otherwise remain hidden in single-engine workflows.
Without such comparison, the Backtrader percabs bug would produce a plausible but incorrect equity curve, the Zipline-Reloaded calendar failure would block execution entirely, and the NautilusTrader issues would either silently truncate results or crash at scale.
Each of these failure modes was discovered only because the affected engine's output was compared against the consensus of the retained engines.

%% file: chapters/07-practical-implications.tex
\section{Practical Implications}\label{sec:implications}

The preceding results converge on a single message: implementation risk is real, measurable, and structured.
This section translates those findings into actionable guidance.

\subsection{When does engine choice matter?}\label{sec:implications:when}

The answer has three tiers.
For simple allocation rules and signal-driven strategies (12 of the 15 benchmarks), the maximum pairwise divergence in annualised Sharpe ratio is below 0.75\%.
At this level the economic impact is negligible; any of the five engines would yield the same investment conclusion.

For high-turnover or cost-intensive strategies (BM03, BM04, BM11), divergence rises to the 2--4\% range and becomes material.
A 3.71\% divergence in Sharpe ratio (BM04) is large enough to shift a strategy's ranking in a multi-strategy selection exercise or alter a risk-budgeting model's output.

The cost-intensity gradient (Section~\ref{sec:results}) provides a predictive rule: strategies characterised by high turnover and high proportional cost rates are precisely those that require multi-engine validation.
Practitioners can triage validation effort accordingly: run a second engine only when the cost-intensity score exceeds a risk-tolerance-calibrated threshold.

\begin{figure}[htbp]
\centering
\includegraphics[width=0.9\textwidth]{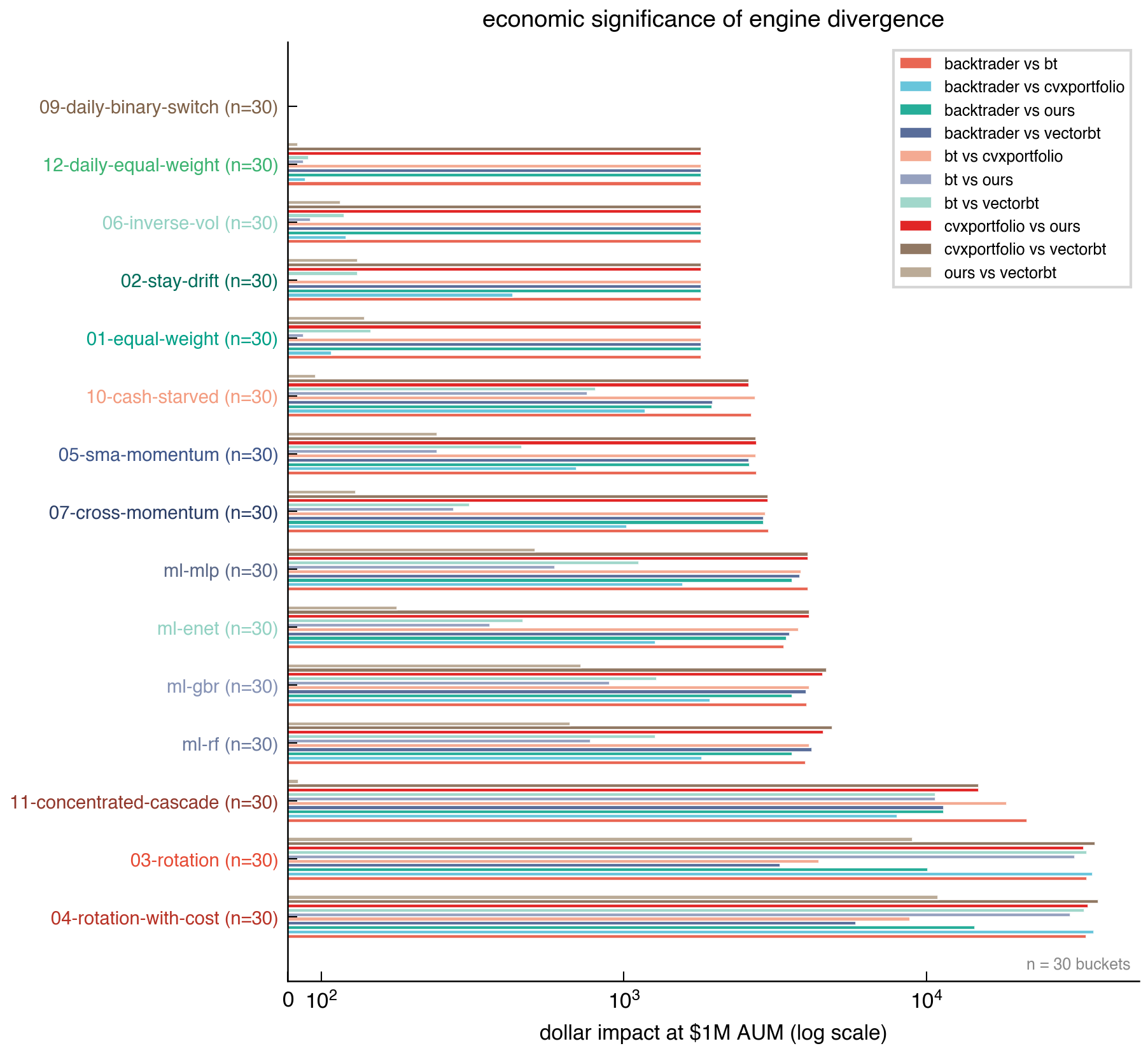}
\caption{Dollar-value translation of implementation risk. Each bar shows the implied annual ambiguity per \$1\,B \acs{aum} for a given benchmark, computed as max pairwise divergence in annualised return.}\label{fig:economic}
\end{figure}

A 0.10\% divergence in annualised return corresponds to approximately \$1\,M per year in ambiguity for a portfolio with \$1\,B in \ac{aum}.
This linear translation lets institutions scale the figure to their own asset base (Figure~\ref{fig:economic}).

For the worst-case benchmark (BM04, 3.71\% divergence), the implied ambiguity is approximately \$37\,M per year per \$1\,B \acs{aum}, material whenever the margin between competing strategies is thin or capital is allocated by Sharpe-ratio ranking.

\begin{figure}[htbp]
\centering
\includegraphics[width=\textwidth]{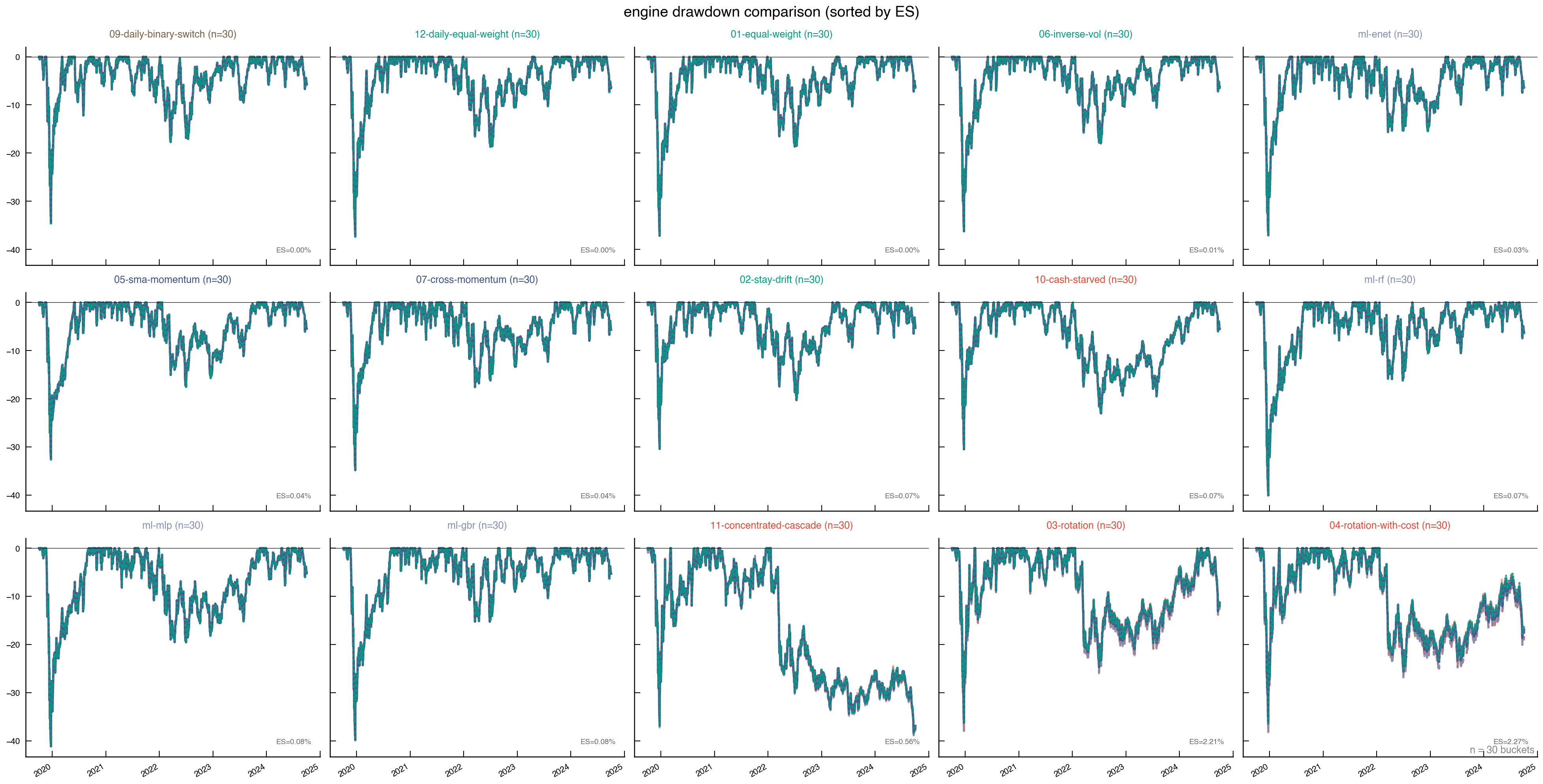}
\caption{Drawdown comparison across engines for all 15 benchmarks. Differences in drawdown statistics are concentrated in rotation and high-turnover strategies.}\label{fig:drawdown-comparison}
\end{figure}

Maximum drawdown, a metric of particular concern to risk managers, also exhibits engine-dependent variation. Figure~\ref{fig:drawdown-comparison} compares drawdown profiles across engines for each benchmark.
For simple and signal strategies the drawdown paths are virtually indistinguishable, whereas rotation benchmarks (BM03, BM04) show drawdown discrepancies that mirror the return divergences documented in Table~\ref{tab:divergence}: the same cost-model timing conventions that shift total returns also shift the depth of measured drawdowns.

\begin{figure}[htbp]
\centering
\includegraphics[width=\textwidth]{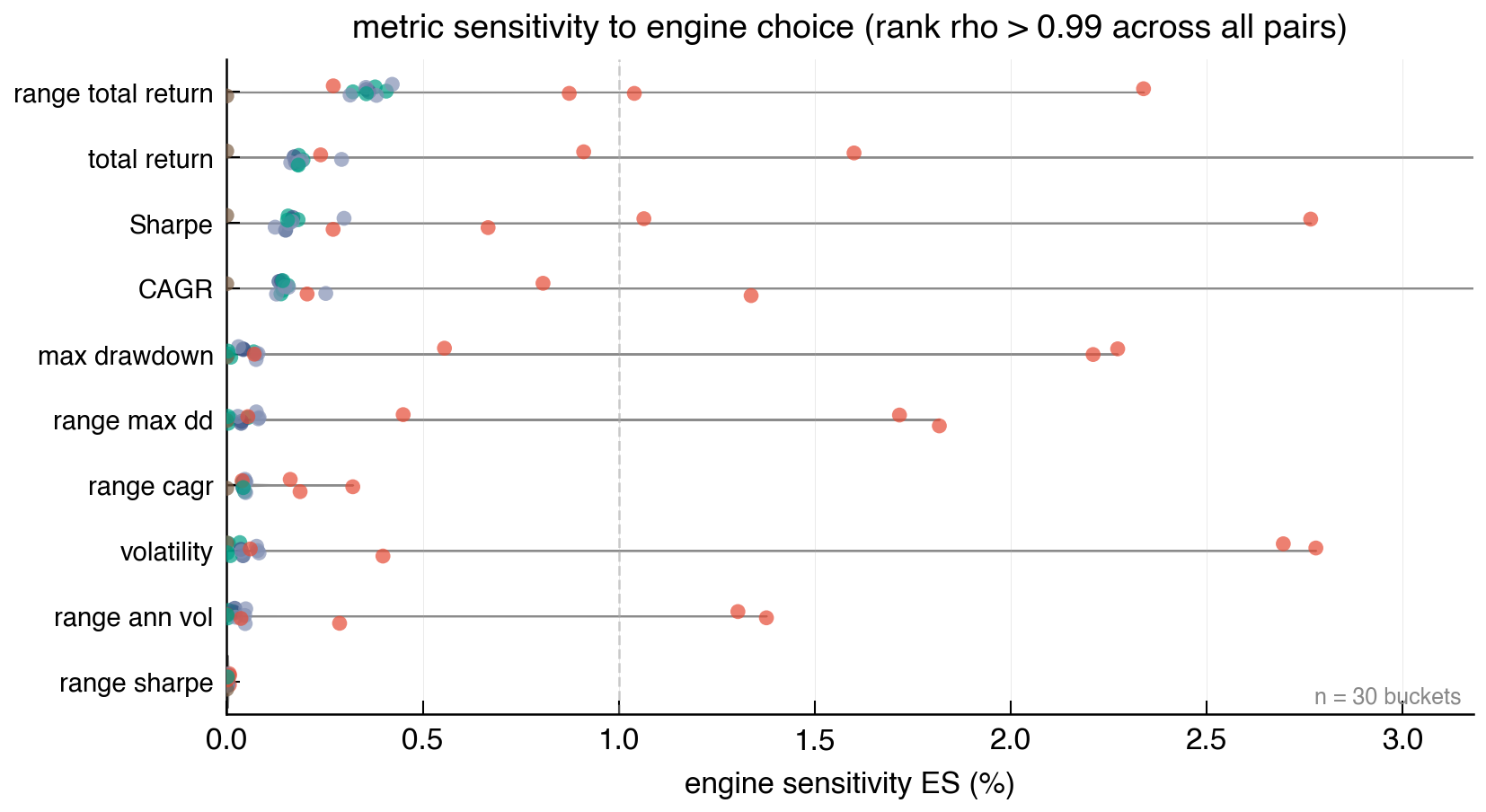}
\caption{Sensitivity of standard performance metrics to engine choice. Each dot represents one benchmark; the horizontal axis measures the \acs{es} in percentage points. Most benchmarks cluster near zero; outliers belong to rotation strategies. Rank correlation exceeds 0.99 across all engine pairs for every metric, which indicates that ordinal conclusions are robust even where cardinal values diverge.}\label{fig:metric-sensitivity}
\end{figure}

Figure~\ref{fig:metric-sensitivity} extends this comparison to all standard performance metrics. for 12 of the 15~benchmarks, the engine spread remains below 0.5\% on every metric, and ordinal rankings are preserved (rank $\rho > 0.99$ for every pair). the outliers belong exclusively to rotation strategies, where Sharpe ratio, max drawdown, and annualised volatility each show engine spreads exceeding 2\%. CAGR shows intermediate sensitivity. the within-bucket range of Sharpe ratios, by contrast, is nearly unaffected by engine choice across all benchmarks, because the cost-model offset shifts all buckets' Sharpe ratios by approximately the same amount and thus preserves cross-bucket dispersion. this metric-level decomposition lets practitioners identify which summary statistics require multi-engine verification for a given strategy type.

\subsection{Multi-validator recommendation}\label{sec:implications:multi}

\begin{figure}[htbp]
\centering
\includegraphics[width=\textwidth]{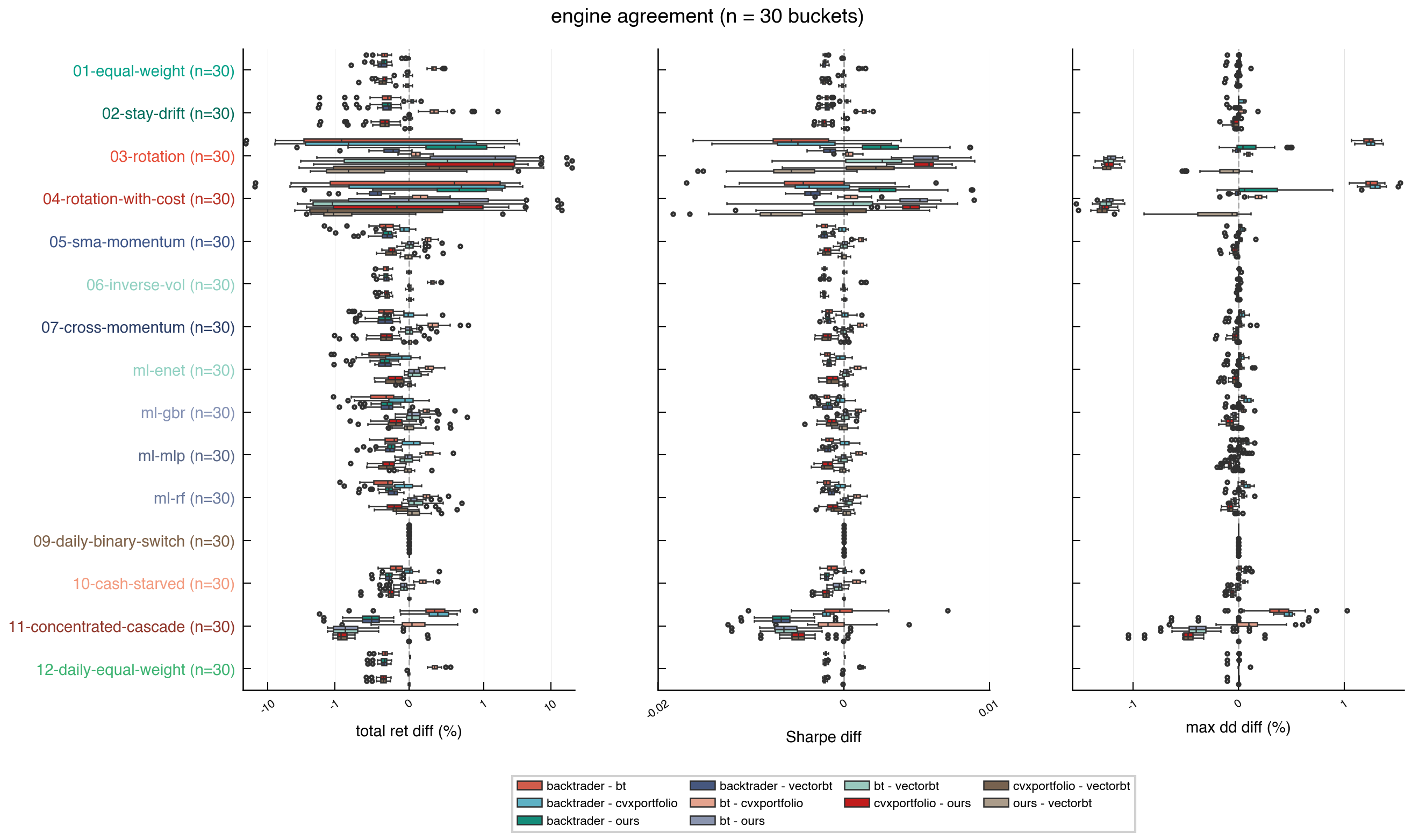}
\caption{Pairwise engine agreement across all 15 benchmarks. Each cell shows the mean absolute divergence for one engine pair and one benchmark. The pattern indicates that agreement depends on cost intensity rather than on any specific engine pair.}\label{fig:agreement}
\end{figure}

A natural response to implementation risk is to validate one's backtest against a second engine.
Our results suggest that a single alternative validator may not suffice.
Pairwise divergence profiles (Section~\ref{sec:results}, Figure~\ref{fig:agreement}) are not perfectly correlated across engine pairs: two engines may agree on one strategy family yet diverge on another, so validating against one alternative provides no protection against a failure mode specific to a third.

We recommend at least two independent validators, chosen to maximise implementation diversity (e.g.\ one event-driven and one vectorised engine), with an explicit audit of each engine's cost model against a reference specification such as Algorithm~\ref{alg:backtest}.
The cost model is the single component driving divergence; verifying correct parameterisation eliminates the dominant source of implementation risk.

The three excluded engines illustrate what happens when implementation risk goes undetected.
Zipline-Reloaded's calendar bug prevented execution entirely, NautilusTrader's compounding defects would have silently corrupted a live portfolio, and QuantConnect Lean's C\# core made faithful strategy translation impossible.
Even among retained engines, Backtrader's \texttt{percabs} default would have produced a 100$\times$ cost undercharge, which would have yielded plausible but incorrect equity curves had it not been caught during integration.
Each of these failures was discovered only through systematic multi-engine comparison, which validates the central recommendation of this section.

\subsection{Connection to backtest overfitting}\label{sec:implications:overfitting}

Implementation risk and statistical overfitting are orthogonal sources of backtest unreliability (Section~\ref{sec:related}).
A strategy that survives deflated Sharpe ratio correction may still produce different economics in a different engine; conversely, perfect cross-engine agreement does not guard against in-sample overfitting.
A complete validation framework therefore requires both: multiple-testing adjustment (e.g.\ the deflated Sharpe ratio of \citet{bailey_deflated_2014}) for the strategy-selection layer, and multi-engine comparison for the simulation layer.

%% file: chapters/08-conclusion.tex
\section{Conclusion}\label{sec:conclusion}

This paper introduced implementation risk, the variability in backtest outcomes attributable solely to engine choice, and provided the first large-scale measurement of its magnitude.
Causal isolation is anchored by a reference implementation developed for this study whose cost model is a direct translation of the proportional-cost specification (Algorithm~\ref{alg:backtest}), which provides a ground-truth baseline against which all divergence is measured.

At zero cost all five engines produce identical equity curves, which establishes perfect causal isolation of the cost-model channel.
Under nonzero costs, divergence scales monotonically with cost intensity (Spearman $\rho = 0.93$, $p < 0.001$) and partitions into three tiers: negligible ($< 0.75\%$, 12 of 15 benchmarks), moderate ($\approx 2\%$, BM11), and material (3.6--3.7\%, the two full-rotation strategies BM03 and BM04).
$\CSI = 0$ for all 15 strategies; engine choice never reverses the sign of the Sharpe ratio.
Forensic analysis of three engines examined during quality control revealed seven distinct defects spanning all five failure-mode categories, with cost-model bugs as the most frequent.

For most strategies engine choice is a second-order concern, but for cost-intensive strategies the implied ambiguity can reach tens of millions of dollars per \$1\,B \acs{aum}, which warrants routine multi-engine validation.

Taken together, these findings deliver the three contributions outlined in Section~\ref{sec:intro}: (1)~to our knowledge, the first large-scale cross-engine comparison under proportional cost models; (2)~a formal four-metric measurement framework grounded in metrology; and (3)~an open benchmark suite for ongoing community validation.
The practical conclusion is that implementation risk, though modest for most strategies, is neither negligible nor unpredictable: it concentrates in cost-intensive strategies and can be detected with a straightforward multi-engine protocol.

Several limitations remain.
We tested only proportional cost models; fixed-cost, tiered, and market-impact models may differ.
All benchmarks use market orders at the close; limit-order and intraday strategies add execution-model complexity not yet addressed.
Five engines, though the broadest comparison in the literature, do not exhaust the available universe.
The zero \acs{csi} may partly reflect large absolute Sharpe ratios; strategies near the zero-Sharpe boundary deserve separate study.
of these, we consider the restriction to proportional costs the most pressing to address next: real-world trading costs are predominantly nonlinear, and since the cost channel is where divergence concentrates, the practical importance of multi-engine validation may be larger than our current estimates suggest.

From a regulatory perspective, our findings suggest that SR~11-7 model validation should be extended to cover the simulation layer, where \acs{es} and \acs{csi} serve as natural acceptance metrics.

We release the full benchmark suite as an open resource; we intend it to serve as a correctness counterpart to MLPerf's speed benchmarks \citep{mattson_mlperf_2020}: a common reference point for certifying engine fidelity and quantifying residual implementation uncertainty.

Future work should examine rolling-window divergence for temporal stability and threshold-free cluster enhancement for robustness across the full parameter space of cost rates and rebalancing frequencies.

%% file: chapters/A1-algorithm.tex
\section{Reference Implementation Pseudocode}\label{sec:algorithm}

Algorithm~\ref{alg:backtest} presents the reference proportional-cost backtest loop that serves as the ground-truth specification for our benchmark suite.
All five retained engines are expected to produce identical equity curves when their cost models faithfully implement this logic, a prediction supported by the zero-cost experiments in Section~\ref{sec:results}.

\begin{algorithm}[htbp]
\caption{Proportional-cost backtest loop.}\label{alg:backtest}
\begin{algorithmic}[1]
\Require weight schedule $W \in \mathbb{R}^{T \times N}$, close prices $P \in \mathbb{R}^{T \times N}$, initial capital $C_0 > 0$, cost rate $c \geq 0$
\Ensure daily equity series $\{E_t\}_{t=1}^{T}$
\State $\mathbf{h} \gets \mathbf{0} \in \mathbb{R}^N$ \Comment{share holdings}
\State $\text{cash} \gets C_0$
\For{each rebalance date $t = 1, \dots, T$}
    \State $V_t \gets \text{cash} + \mathbf{h}^\top \mathbf{p}_t$ \Comment{mark to market}
    \State $\mathbf{w}_t \gets W_{t,:}$ \Comment{target weights from schedule}
    \State $\boldsymbol{\delta} \gets \mathbf{w}_t \cdot V_t - \mathbf{h} \odot \mathbf{p}_t$ \Comment{trade deltas}
    \State $\text{cost}_t \gets c \cdot \lVert \boldsymbol{\delta} \rVert_1$ \Comment{proportional transaction cost}
    \State $V_t^{\text{net}} \gets V_t - \text{cost}_t$ \Comment{post-cost portfolio value}
    \State $\mathbf{h} \gets (\mathbf{w}_t \cdot V_t^{\text{net}}) \oslash \mathbf{p}_t$ \Comment{reallocate to target weights}
    \State $\text{cash} \gets V_t^{\text{net}} - \mathbf{h}^\top \mathbf{p}_t$ \Comment{residual cash}
    \State $E_t \gets V_t^{\text{net}}$
\EndFor
\State \Return $\{E_t\}_{t=1}^{T}$
\end{algorithmic}
\end{algorithm}

The algorithm operates in three phases per rebalance date.
First, it marks the portfolio to market by computing the current value $V_t$ from share holdings and close prices (line~4).
Second, it computes trade deltas $\boldsymbol{\delta}$ as the vector difference between target and current dollar positions, then deducts the proportional cost $c \cdot \lVert \boldsymbol{\delta} \rVert_1$ from the portfolio value (lines~6--8).
Third, it reallocates the post-cost value $V_t^{\text{net}}$ according to the target weights and updates share holdings via element-wise division by the price vector (lines~9--10).

Two implementation details deserve comment.
The cost is deducted from the total portfolio value before reallocation, which means that the cost reduces all positions proportionally rather than being charged against cash alone.
This ``cost-then-allocate'' convention is the simplest correct approach and matches the default behaviour of most vectorised engines.
Event-driven engines that deduct cost on a per-order basis must produce the same aggregate deduction for the algorithm to be considered faithful.

For strategies that rebalance less frequently than daily, the equity series on non-rebalance days is computed by mark-to-market: $E_t = \text{cash} + \mathbf{h}^\top \mathbf{p}_t$ with no trading.
The algorithm as written shows only the rebalance-day logic; the daily mark-to-market pass is implicit.

The second detail is that the algorithm assumes fractional shares.
All five retained engines are configured to allow fractional holdings, which eliminates rounding as a source of divergence.
The impact of integer-share constraints is left to future work.

%% file: chapters/A2-engine-details.tex
\section{Engine Configuration Details}\label{sec:engines}

Table~\ref{tab:engines} lists the five retained engines together with their versions, Python packages, and configuration parameters.
Each engine was configured to match the reference specification in Algorithm~\ref{alg:backtest} as closely as the \acs{api} permits; non-obvious settings are noted below.

\begin{table}[htbp]
\caption{Engine configurations used in the benchmark suite.}\label{tab:engines}
\begin{tabularx}{\textwidth}{lllX}
\toprule
Engine & Version & Package & Configuration notes \\
\midrule
Ours & -- & (internal) & direct translation of Algorithm~\ref{alg:backtest}; fractional shares; cost deducted before reallocation \\
bt & 1.1.0 & \texttt{bt} & \texttt{bt.Backtest} with commissions lambda; \texttt{integer\_positions=False} \\
vectorbt & 0.26.2 & \texttt{vectorbt} & \texttt{Portfolio.from\_orders} with \texttt{size\_type="targetpercent"}, \texttt{group\_by=True}, \texttt{cash\_sharing=True} \\
Backtrader & 1.9.78 & \texttt{backtrader} & \texttt{percabs=True} required; sells-first ordering fix applied \\
cvxportfolio & 1.3.0 & \texttt{cvxportfolio} & forward returns $r_t = (p_{t+1} - p_t) / p_t$ with \texttt{shift(-1)} to align close-to-close convention \\
\botrule
\end{tabularx}
\end{table}

\subsection{Our engine}
The reference implementation is a direct translation of Algorithm~\ref{alg:backtest} into NumPy vectorised operations.
It was developed specifically for this study to serve as a minimal, auditable reference specification, rather than adopted from a pre-existing library.
Divergence measured against it reflects disagreement with this specification, not absolute error in any external sense.
It accepts a weight matrix $W$, a close-price matrix $P$, initial capital $C_0$, and a scalar cost rate $c$, and returns the daily equity series.
No execution model, slippage model, or calendar logic is involved; the implementation is intentionally minimal so that the code and the specification are the same thing.

\subsection{bt (pmorissette)}
bt provides a tree-structured strategy \acs{api} in which each node represents an allocation step.
Commissions are specified as a callable that receives the dollar trade value and returns the fee.
We set \texttt{integer\_positions=False} to enable fractional shares and passed a lambda that returns $|q| \cdot p \cdot c$ as the proportional cost for each fill.

\subsection{vectorbt}
vectorbt is a fully vectorised engine that operates on NumPy arrays rather than simulating order-by-order.
The \texttt{Portfolio.from\_orders} constructor accepts target-percent orders directly via \texttt{size\_type="targetpercent"}.
We enabled \texttt{group\_by=True} and \texttt{cash\_sharing=True} so that all assets within a benchmark share a single cash account, which matches the single-portfolio assumption of Algorithm~\ref{alg:backtest}.

\subsection{Backtrader}
Backtrader is an event-driven engine that processes bars sequentially.
Two non-obvious settings were required.
First, \texttt{percabs=True} must be set explicitly to prevent the engine from dividing the commission rate by~100 (Section~\ref{sec:forensic:bt}).
Second, the fill-ordering logic was patched to process sell orders before buy orders, which prevents spurious margin rejections during rebalancing.
Both issues are documented in detail in the forensic analysis (Section~\ref{sec:forensic}).

\subsection{cvxportfolio}
cvxportfolio models returns as forward-looking: $r_t = (p_{t+1} - p_t) / p_t$.
To align with the close-to-close convention used by the other engines, the return series must be shifted by one period with \texttt{shift(-1)}.
Without this adjustment the engine's equity curve is offset by one day, which produces apparent divergence that is purely a timing artefact rather than a genuine cost-model difference.

%% file: chapters/A3-statistical-diagnostics.tex
\section{Statistical Diagnostics}\label{sec:diagnostics}

This appendix collects diagnostic analyses that validate the statistical methodology used in the main text.

\subsection{Normality assessment}

\begin{figure}[htbp]
\centering
\includegraphics[width=\textwidth]{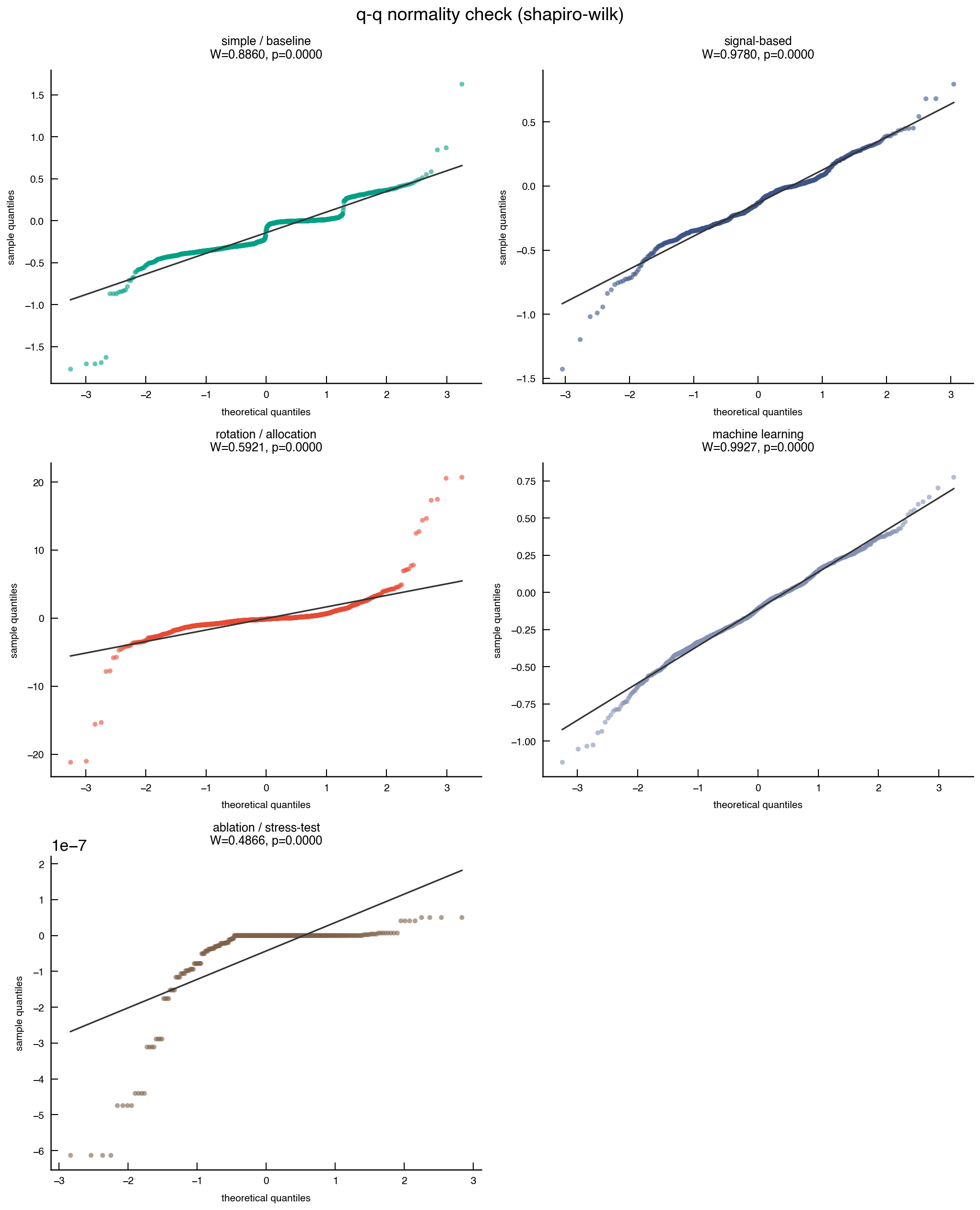}
\caption{Q--Q plots of per-bucket divergence for each benchmark category. Departures from the diagonal indicate non-normality; the rotation category shows the largest tails.}\label{fig:qq-normality}
\end{figure}

The primary $t$-test of mean pairwise divergence assumes approximately normal distributions within each benchmark category.
Figure~\ref{fig:qq-normality} shows Q--Q plots for the per-bucket divergence series, stratified by category.
The simple, signal, and ablation categories track the normal reference line closely; the \acs{ml} category shows minor departures in the tails; the rotation category exhibits heavier tails, consistent with the higher cross-bucket variance documented in Section~\ref{sec:results}.
These departures motivate the non-parametric robustness checks reported below.

\subsection{Power analysis}

\begin{figure}[htbp]
\centering
\includegraphics[width=0.55\textwidth]{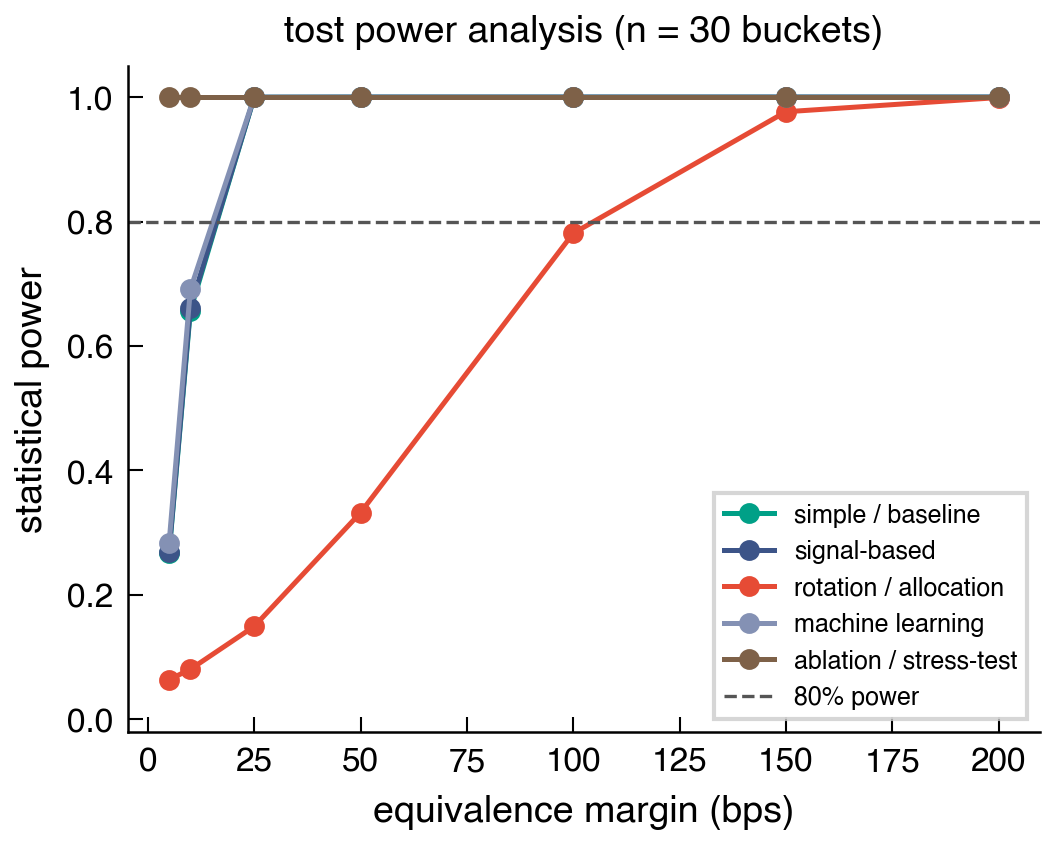}
\caption{Statistical power versus equivalence margin by benchmark category. The $n = 30$ design achieves greater than 80\% power at 10--25~bps for simple, signal, \acs{ml}, and ablation categories; rotation strategies require margins of 100--130~bps to reach 80\% power.}\label{fig:power-analysis}
\end{figure}

For simple, signal, \acs{ml}, and ablation benchmarks, $n = 30$ buckets provides greater than 80\% power at equivalence margins of 10--25~bps (Figure~\ref{fig:power-analysis}).
For rotation strategies (BM03, BM04, BM10, BM11), power reaches 80\% only at margins of approximately 100--130~bps, which reflects the higher cross-bucket variance in this category.

\subsection{Permutation test distributions}

\begin{figure}[htbp]
\centering
\includegraphics[width=\textwidth]{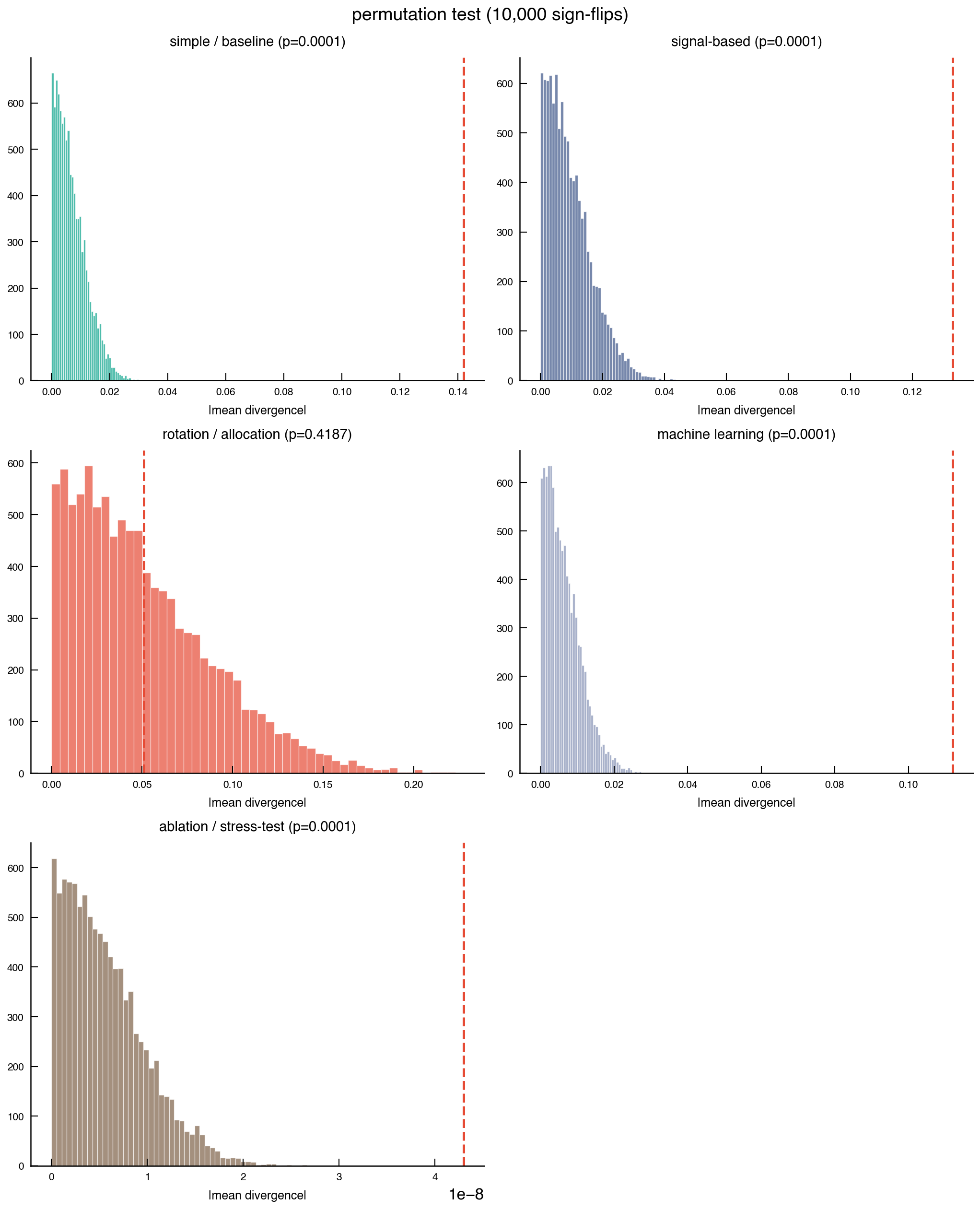}
\caption{Monte Carlo permutation test null distributions. The observed mean absolute divergence (vertical line) falls far outside the null distribution for simple, signal, \acs{ml}, and ablation categories ($p = 0.0001$). Rotation is non-significant ($p = 0.4187$) due to high cross-bucket variance.}\label{fig:permutation-test}
\end{figure}

Figure~\ref{fig:permutation-test} shows the Monte Carlo null distributions (10{,}000 draws) for each benchmark category alongside the observed test statistic.
The simple, signal, \acs{ml}, and ablation categories reject the null of no divergence ($p = 0.0001$). The rotation category does not ($p = 0.4187$); this non-rejection reflects high cross-bucket variance rather than absence of divergence.

\subsection{Wilcoxon robustness}

\begin{figure}[htbp]
\centering
\includegraphics[width=0.7\textwidth]{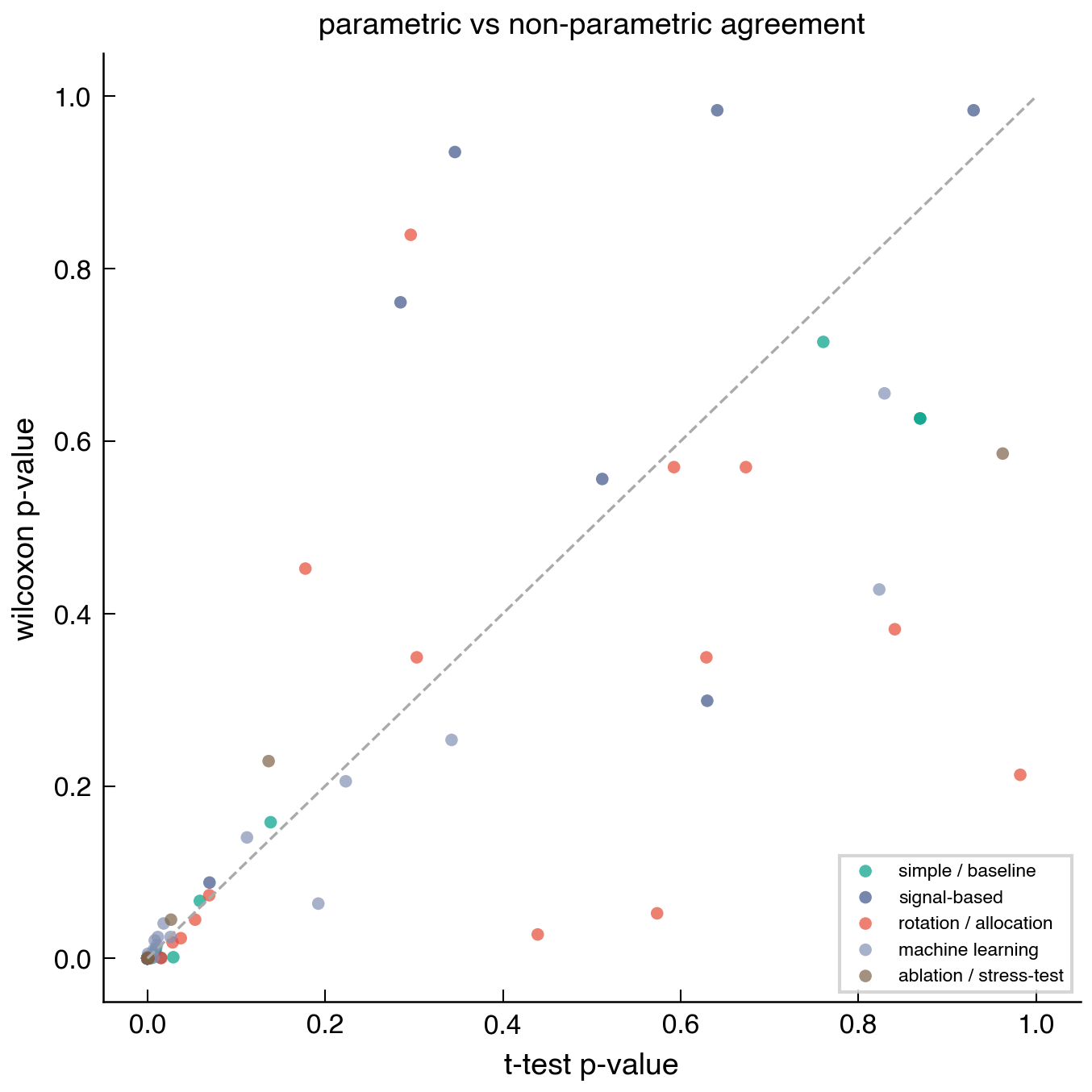}
\caption{Wilcoxon signed-rank $p$-values versus parametric $t$-test $p$-values for all benchmark-pair combinations. Points near the diagonal indicate agreement between parametric and non-parametric tests.}\label{fig:wilcoxon-robustness}
\end{figure}

The non-parametric Wilcoxon signed-rank test provides a distribution-free check on the parametric $t$-test conclusions. Figure~\ref{fig:wilcoxon-robustness} compares $p$-values from both tests for every benchmark-pair combination. All points cluster near the diagonal, which indicates that the two tests agree and that the parametric conclusions are not artefacts of distributional assumptions.

%% file: chapters/A4-data-quality.tex
\section{Data Quality Control}\label{sec:dataqc}

This appendix documents the quality-control checks applied to the price data and the stratified bucket construction.

\subsection{Price data}

\begin{figure}[htbp]
\centering
\includegraphics[width=\textwidth]{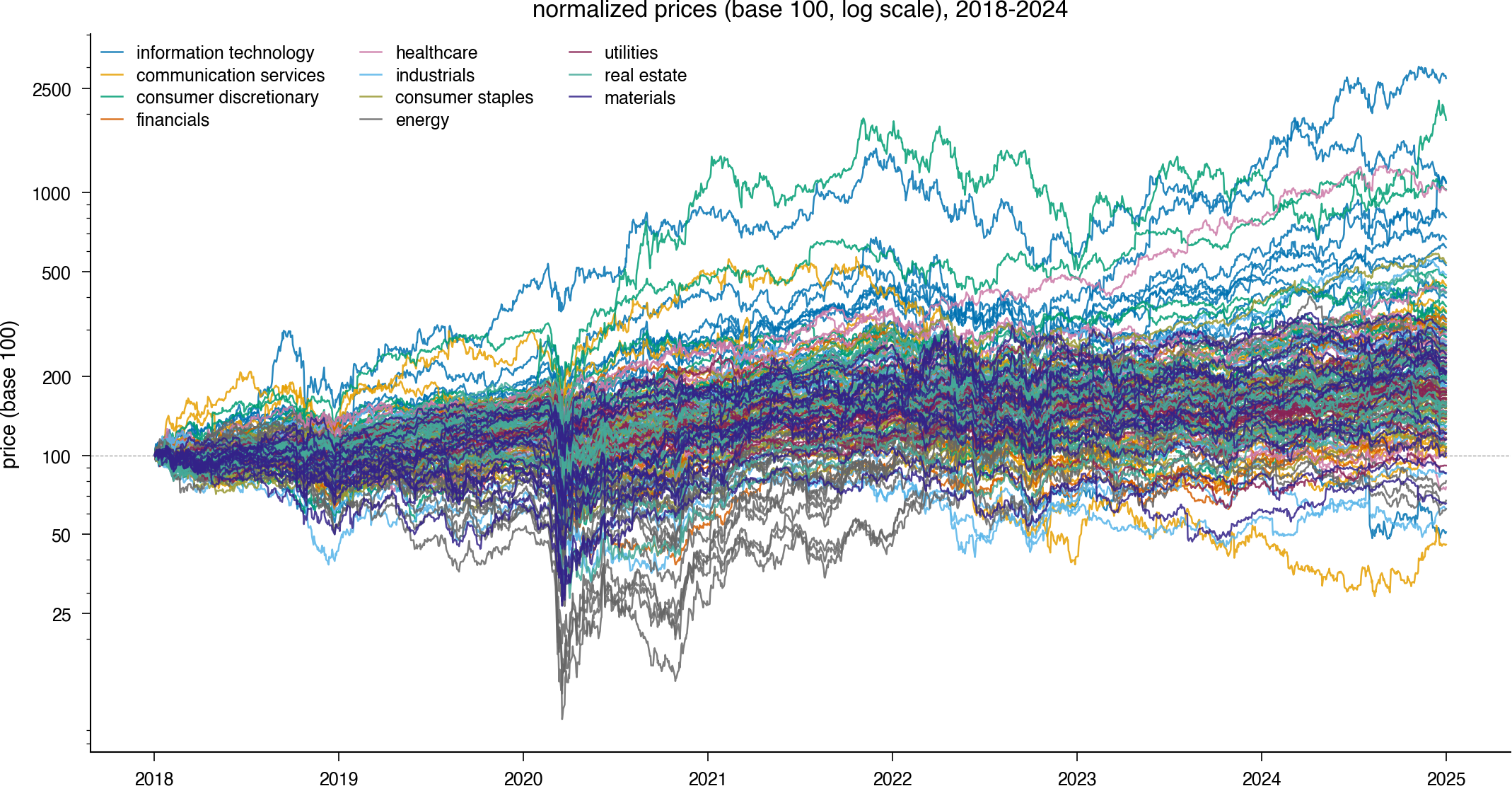}
\caption{Log-scale adjusted-close prices for all 180 stocks over the full download period (2018--2024), coloured by sector. Portfolio evaluation begins in January~2020; the preceding two years provide the \acs{ml} training and signal warm-up buffer described in the text. No gaps or stale prices are present.}\label{fig:qc-prices}
\end{figure}

The full downloaded dataset comprises 180 S\&P~500 constituents with complete daily adjusted-close histories from January~2018 to December~2024 (1{,}761 trading days, zero missing values).
The first two years (2018--2019) serve as a look-back buffer for \acs{ml} training and signal warm-up; all portfolio evaluation uses the backtest period of January~2020 to December~2024 (1{,}258 trading days).
Only adjusted-close prices are used; no open, high, low, or volume data enter any strategy or engine.
Figure~\ref{fig:qc-prices} plots log-scale, sector-coloured price trajectories for the full seven-year download, which includes the training buffer.
The backtest period spans the COVID-19 drawdown (February to March 2020), the 2022 rate-hiking sell-off, and the subsequent recovery, which provides a mix of bull, bear, and range-bound regimes.

All 180 stocks have complete daily observations for every trading day in the sample; no stock-month pair contains missing data and no imputation was required.
Table~\ref{tab:desc-stats} summarises the universe-level and sector-level descriptive statistics over the backtest period.

\begin{table}[htbp]
\centering
\footnotesize
\caption{descriptive statistics for the 180-stock universe, computed over the backtest period (January~2020 to December~2024). panel~A reports aggregate statistics across all stocks. panel~B reports sector-level averages of annualised volatility, annualised return, and maximum drawdown.}\label{tab:desc-stats}
\vspace{4pt}
\begin{tabular}{@{}lr@{}}
\toprule
\multicolumn{2}{l}{\textit{panel A: universe-level statistics}} \\
\midrule
number of stocks        & 180 \\
full download period    & Jan 2018 to Dec 2024 (1{,}761 days) \\
backtest period         & Jan 2020 to Dec 2024 (1{,}258 days) \\
mean annualised return (\%)     & 19.36 \\
mean annualised volatility (\%) & 33.81 \\
mean maximum drawdown (\%)      & 47.59 \\
mean pairwise correlation       & 0.39 \\
min / max pairwise correlation  & $-$0.05\,/\,0.92 \\
\botrule
\end{tabular}

\vspace{6pt}

\begin{tabular}{@{}lrrrr@{}}
\toprule
\multicolumn{5}{l}{\textit{panel B: sector breakdown}} \\
\midrule
sector & $n$ & vol.\ (\%) & ret.\ (\%) & MDD (\%) \\
\midrule
communication services  & 16 & 35.77 & 10.34 & 56.0 \\
consumer discretionary  & 17 & 36.62 & 36.18 & 50.4 \\
consumer staples        & 16 & 24.69 & 12.55 & 34.5 \\
energy                  & 16 & 45.92 & 15.82 & 65.7 \\
financials              & 16 & 34.62 & 16.53 & 47.8 \\
healthcare              & 16 & 27.94 & 15.43 & 38.7 \\
industrials             & 16 & 31.74 & 15.80 & 45.6 \\
information technology  & 17 & 39.31 & 57.43 & 48.4 \\
materials               & 16 & 35.69 & 13.86 & 48.7 \\
real estate             & 17 & 32.90 & 11.45 & 49.1 \\
utilities               & 17 & 26.75 &  5.48 & 38.9 \\
\botrule
\end{tabular}
\end{table}

\subsection{Bucket stratification}

The 30 non-overlapping buckets of six stocks each are constructed via Mahalanobis rerandomisation \citep{morgan_rerandomization_2012} over three covariates: mean annualised volatility, mean pairwise correlation, and log total return.
From 20{,}000 candidate permutations the procedure selects the assignment that minimises the Mahalanobis distance between bucket-level and population covariate means (best score 43.79 out of 20{,}000 candidates).

\begin{figure}[htbp]
\centering
\includegraphics[width=0.85\textwidth]{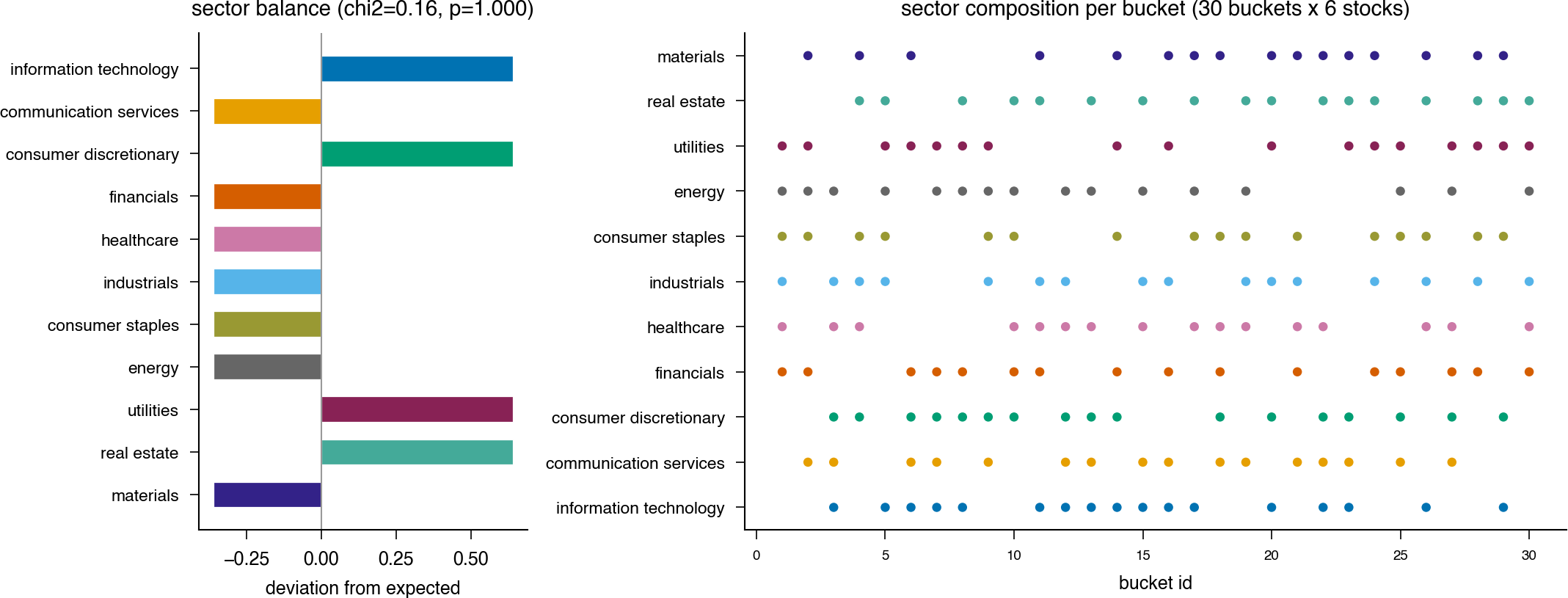}
\caption{Sector representation across the 30 stratified buckets. Each bucket contains six stocks from six distinct sectors ($\chi^2 = 0.156$, $p = 1.00$).}\label{fig:qc-sector}
\end{figure}

A chi-square test of sector balance yields $\chi^2 = 0.156$ ($p = 1.00$) and normalised Shannon entropy of 0.9998, both of which indicate near-perfect sector balance (Figure~\ref{fig:qc-sector}).
Every bucket contains stocks from exactly six distinct sectors.

\begin{figure}[htbp]
\centering
\includegraphics[width=0.7\textwidth]{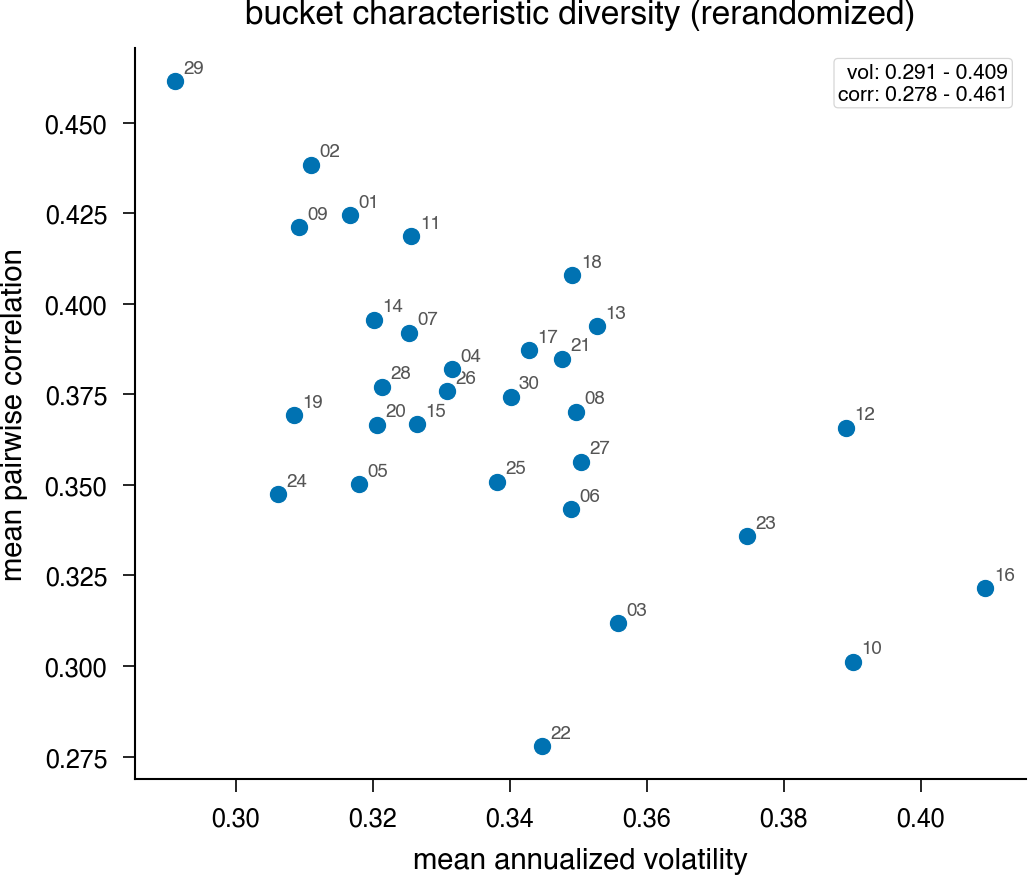}
\caption{Bucket diversity: mean pairwise correlation versus mean annualised volatility for each of the 30 buckets. The spread indicates that the stratification covers a wide range of dependence and risk conditions.}\label{fig:qc-diversity}
\end{figure}

Figure~\ref{fig:qc-diversity} shows that the buckets span mean pairwise correlations of 0.28 to 0.46 and annualised volatilities of 0.29 to 0.41, which ensures that results are not artefacts of a single dependence or risk regime.

%% file: main.bib
@techreport{sr117_2011,
  author       = {{Board of Governors of the Federal Reserve System}},
  title        = {Supervisory Guidance on Model Risk Management},
  number       = {SR Letter 11-7},
  institution  = {Board of Governors of the Federal Reserve System},
  year         = {2011},
  month        = apr
}

@article{perignon_computational_2024,
  author  = {P{\'e}rignon, Christophe and Akmansoy, Olivier and Hurlin, Christophe and Coussin, Anna and Desagr{\'e}, C{\'e}line and Ducret, Tanguy and Journe, Eric},
  title   = {Computational Reproducibility in Finance: Evidence from 1,000 Tests},
  journal = {Review of Financial Studies},
  volume  = {37},
  number  = {11},
  pages   = {3558--3593},
  year    = {2024},
  doi     = {10.1093/rfs/hhae029}
}

@article{alonso_model_2022,
  author  = {Alonso Robisco, Andr{\'e}s and Carb{\'o} Mart{\'i}nez, Jos{\'e} Manuel},
  title   = {Measuring the model risk-adjusted performance of machine learning algorithms in credit default prediction},
  journal = {Financial Innovation},
  volume  = {8},
  pages   = {70},
  year    = {2022},
  doi     = {10.1186/s40854-022-00366-1}
}

@article{giudici_safe_2023,
  author  = {Giudici, Paolo and Raffinetti, Emanuela},
  title   = {{SAFE} artificial intelligence in finance},
  journal = {Finance Research Letters},
  volume  = {56},
  pages   = {104088},
  year    = {2023}
}

@article{bailey_deflated_2014,
  author  = {Bailey, David H. and {Lopez de Prado}, Marcos},
  title   = {The Deflated {S}harpe Ratio: Correcting for Selection Bias, Backtest Overfitting, and Non-Normality},
  journal = {Journal of Portfolio Management},
  volume  = {40},
  number  = {5},
  pages   = {94--107},
  year    = {2014}
}

@article{bailey_probability_2014,
  author  = {Bailey, David H. and Borwein, Jonathan M. and {Lopez de Prado}, Marcos and Zhu, Qiji Jim},
  title   = {Pseudo-Mathematics and Financial Charlatanism: The Effects of Backtest Overfitting on Out-of-Sample Performance},
  journal = {Notices of the American Mathematical Society},
  volume  = {61},
  number  = {5},
  pages   = {458--471},
  year    = {2014}
}

@article{white_reality_2000,
  author  = {White, Halbert},
  title   = {A Reality Check for Data Snooping},
  journal = {Econometrica},
  volume  = {68},
  number  = {5},
  pages   = {1097--1126},
  year    = {2000}
}

@article{hansen_test_2005,
  author  = {Hansen, Peter Reinhard},
  title   = {A Test for Superior Predictive Ability},
  journal = {Journal of Business and Economic Statistics},
  volume  = {23},
  number  = {4},
  pages   = {365--380},
  year    = {2005}
}

@article{romano_stepwise_2005,
  author  = {Romano, Joseph P. and Wolf, Michael},
  title   = {Stepwise Multiple Testing as Formalized Data Snooping},
  journal = {Econometrica},
  volume  = {73},
  number  = {4},
  pages   = {1237--1282},
  year    = {2005}
}

@article{harvey_and_2016,
  author  = {Harvey, Campbell R. and Liu, Yan and Zhu, Heqing},
  title   = {\ldots and the Cross-Section of Expected Returns},
  journal = {Review of Financial Studies},
  volume  = {29},
  number  = {1},
  pages   = {5--68},
  year    = {2016}
}

@article{wiecki_all_2016,
  author  = {Wiecki, Thomas and Campbell, Andrew and Lent, Justin and Stauth, Jessica},
  title   = {All That Glitters Is Not Gold: Comparing Backtest and Out-of-Sample Performance on a Large Cohort of Trading Algorithms},
  journal = {Journal of Investing},
  volume  = {25},
  number  = {3},
  pages   = {69--80},
  year    = {2016}
}

@article{boyd_multiperiod_2017,
  author  = {Boyd, Stephen and Busseti, Enzo and Diamond, Steven and Kahn, Ronald N. and Koh, Kwangmoo and Nystrup, Peter and Speth, Jan},
  title   = {Multi-Period Trading via Convex Optimization},
  journal = {Foundations and Trends in Optimization},
  volume  = {3},
  number  = {1},
  pages   = {1--76},
  year    = {2017}
}

@article{demiguel_optimal_2009,
  author  = {DeMiguel, Victor and Garlappi, Lorenzo and Uppal, Raman},
  title   = {Optimal Versus Naive Diversification: How Inefficient is the {1/N} Portfolio Strategy?},
  journal = {Review of Financial Studies},
  volume  = {22},
  number  = {5},
  pages   = {1915--1953},
  year    = {2009}
}

@article{tebaldi_use_2007,
  author  = {Tebaldi, Claudia and Knutti, Reto},
  title   = {The use of the multi-model ensemble in probabilistic climate projections},
  journal = {Philosophical Transactions of the Royal Society A},
  volume  = {365},
  number  = {1857},
  pages   = {2053--2075},
  year    = {2007}
}

@techreport{jcgm_gum_2008,
  author      = {{JCGM}},
  title       = {Evaluation of measurement data -- Guide to the expression of uncertainty in measurement},
  number      = {JCGM 100:2008},
  institution = {Joint Committee for Guides in Metrology},
  year        = {2008}
}

@book{saltelli_global_2008,
  author    = {Saltelli, Andrea and Ratto, Marco and Andres, Terry and Campolongo, Francesca and Cariboni, Jessica and Gatelli, Debora and Saisana, Michaela and Tarantola, Stefano},
  title     = {Global Sensitivity Analysis: The Primer},
  publisher = {John Wiley and Sons},
  year      = {2008}
}

@inproceedings{marx_predictive_2020,
  author    = {Marx, Charles and Calmon, Flavio and Ustun, Berk},
  title     = {Predictive Multiplicity in Classification},
  booktitle = {Proceedings of the 37th International Conference on Machine Learning (ICML)},
  pages     = {6765--6774},
  year      = {2020}
}

@article{morgan_rerandomization_2012,
  author  = {Morgan, Kari Lock and Rubin, Donald B.},
  title   = {Rerandomization to improve covariate balance in experiments},
  journal = {Annals of Statistics},
  volume  = {40},
  number  = {2},
  pages   = {1263--1282},
  year    = {2012}
}

@book{fisher_design_1935,
  author    = {Fisher, Ronald A.},
  title     = {The Design of Experiments},
  publisher = {Oliver and Boyd},
  address   = {Edinburgh},
  year      = {1935}
}

@article{benjamini_control_2001,
  author  = {Benjamini, Yoav and Yekutieli, Daniel},
  title   = {The control of the false discovery rate in multiple testing under dependency},
  journal = {Annals of Statistics},
  volume  = {29},
  number  = {4},
  pages   = {1165--1188},
  year    = {2001}
}

@article{lin_concordance_1989,
  author  = {Lin, Lawrence I-Kuei},
  title   = {A concordance correlation coefficient to evaluate reproducibility},
  journal = {Biometrics},
  volume  = {45},
  number  = {1},
  pages   = {255--268},
  year    = {1989}
}

@article{schuirmann_comparison_1987,
  author  = {Schuirmann, Donald J.},
  title   = {A comparison of the two one-sided tests procedure and the power approach for assessing the equivalence of average bioavailability},
  journal = {Journal of Pharmacokinetics and Biopharmaceutics},
  volume  = {15},
  number  = {6},
  pages   = {657--680},
  year    = {1987}
}

@techreport{iso5725_1994,
  author      = {{ISO}},
  title       = {Accuracy (trueness and precision) of measurement methods and results},
  number      = {ISO 5725-1:1994},
  institution = {International Organization for Standardization},
  year        = {1994}
}

@misc{morissette_bt_2014,
  author  = {Morissette, Philippe},
  title   = {bt: flexible backtesting for {P}ython},
  url     = {https://github.com/pmorissette/bt},
  year    = {2014}
}

@misc{polakow_vectorbt_2020,
  author  = {Polakow, Oleg},
  title   = {vectorbt: vectorised backtesting, analytics, and optimisation for {P}ython},
  url     = {https://github.com/polakowo/vectorbt},
  year    = {2020}
}

@misc{rodriguez_backtrader_2015,
  author  = {Rodriguez, Daniel},
  title   = {backtrader: {P}ython backtesting library for trading strategies},
  url     = {https://github.com/mementum/backtrader},
  year    = {2015}
}

@misc{jansen_zipline_2020,
  author  = {Jansen, Stefan},
  title   = {zipline-reloaded: a {P}ythonic algorithmic trading library},
  url     = {https://github.com/stefan-jansen/zipline-reloaded},
  year    = {2020},
  note    = {Community continuation of Quantopian's Zipline}
}

@misc{nautechsystems_nautilus_2021,
  author  = {{Nautech Systems}},
  title   = {{NautilusTrader}: high-performance algorithmic trading platform},
  url     = {https://github.com/nautechsystems/nautilus_trader},
  year    = {2021}
}

@misc{quantconnect_lean_2013,
  author  = {{QuantConnect}},
  title   = {{Lean}: open-source algorithmic trading engine},
  url     = {https://github.com/QuantConnect/Lean},
  year    = {2013}
}

@inproceedings{mattson_mlperf_2020,
  author    = {Mattson, Peter and Cheng, Christine and Diamos, Gregory and Coleman, Cody and Micikevicius, Paulius and Patterson, David and Tang, Hanlin and Wei, Gu-Yeon and Bailis, Peter and Bittorf, Victor and Brooks, David and Chen, Dehao and Dutta, Debojyoti and Gupta, Udit and Hazelwood, Kim and Hock, Andrew and Huang, Xinyuan and Ike, Atsushi and Jia, Bill and Kang, Daniel and Kanter, David and Kumar, Naveen and Liao, Jeffery and Ma, Guokai and Narayanan, Deepak and Oguntebi, Tayo and Pekhimenko, Gennady and Pentecost, Lillian and Reddi, Vijay Janapa and Robie, Taylor and St John, Tom and Tabber, Tsuguchika and Wu, Carole-Jean and Xu, Lingjie and Yamazaki, Masafumi and Young, Cliff and Zaharia, Matei},
  title     = {{MLPerf} Training Benchmark},
  booktitle = {Proceedings of Machine Learning and Systems (MLSys)},
  year      = {2020}
}

@misc{aroussi_yfinance_2024,
  author  = {Aroussi, Ran},
  title   = {{yfinance}: download market data from {Yahoo!\ Finance's API}},
  url     = {https://github.com/ranaroussi/yfinance},
  year    = {2024},
  note    = {Python package, version 0.2.36}
}
